\documentclass[%
amsmath,amssymb,
aps,
prb,
]{revtex4-1}

\usepackage{graphicx}
\usepackage{amssymb}
\usepackage{amsfonts}
\usepackage{amsmath}
\usepackage{amsbsy}
\usepackage{natbib}
\usepackage{comment}

\usepackage{color}
\usepackage{float}

\renewcommand{\vec}[1]{\mbox{\boldmath$\mathrm{#1}$}}
\let\sb=_ \catcode`\_=\active \def_#1{\ensuremath \sb{\rm#1}}

\begin{document}
	
	
	\title{Electron-magnon spin conversion and magnonic spin pumping in antiferromagnet/heavy metal heterostructure}
	\author{Xi-guang Wang$^{1}$}
	\author{Yao-Zhuang Nie$^{1}$}
	\author{L. Chotorlishvili$^{2}$}
	\author{Qing-lin Xia$^{1}$}
	\author{J. Berakdar$^{2}$}
	\author{Guang-hua Guo$^1$}
	\email{guogh@mail.csu.edu.cn}
	
	\affiliation{
		$^{1}$School of Physics and Electronics, Central South University, Changsha 410083, China \\
		$^{2}$ Institut f\"ur Physik, Martin-Luther Universit\"at Halle-Wittenberg, D-06120 Halle/Saale, Germany
	}
	
	\date{\today}

\begin{abstract}
We study the exchange between electron and magnon spins at the interface of an antiferromagnet and a heavy metal at finite temperatures. 
The underlying physical mechanism is based on  spin torque  associated with the creation/annihilation of thermal magnons with right-hand and left-hand polarization. The creation/annihilation process  depends strongly on the relative orientation between the polarization of the electron and the magnon spins. For a sufficiently  strong  spin transfer torque (STT), the conversion process becomes nonlinear, generating a nonzero net spin pumping current in the AFM that can  detected in the neighboring metal layer. Applying an external magnetic field renders possible the manipulation  of the STT driving thermal spin pumping. Our theoretical results are experimentally feasible and are of  a direct relevance to  antiferromagnet-based spintronic devices.

\end{abstract}

\maketitle

\section{Introduction}
 Electronic and magnonic spin currents are central to spintronics  \cite{RevModPhys.76.323,Cchumak2015,Wolf1488,Ohno2010,RevModPhys.87.1213,PhysRevLett.107.066604}. 
  Electronic spin current is generated for instance due to the spin Hall effect (SHE) in a non-magnetic metal layer\cite{PhysRevLett.83.1834,Valenzuela2006,RevModPhys.87.1213,Kato1910}, or  through the oscillation of magnetization in a ferromagnetic layer (spin pumping) \cite{RevModPhys.77.1375,doi:10.1063/1.2199473,PhysRevLett.106.216601,doi:10.1063/1.3587173}.
The magnonic spin current, meaning a flux of non-equilibrium magnons results from an  applied temperature gradient, microwave field, or due to electronic spin-transfer torque
\cite{Cchumak2015,Uchida2010,doi:10.1063/1.3689787,Kajiwara2010,PhysRevB.95.020414,PhysRevB.99.024410,PhysRevB.86.054445}.
 Spin currents in antiferromagnets (AFMs) are also highly interesting for AFM spintronics \cite{Jungwirth2016,doi:10.1063/1.4862467, RevModPhys.90.015005, Gomonay2018, JUNGFLEISCH2018865}.
  Several spin transport phenomena, including SHE, spin Seebeck effect, N$ {\rm \acute{e}} $el spin-orbit torque were reported and their potential for applications were discussed in AFMs \cite{Jungwirth2016,doi:10.1063/1.4862467,RevModPhys.90.015005,Gomonay2018,JUNGFLEISCH2018865,Park2011,PhysRevB.93.014425,PhysRevB.93.054412,
Wadley587,PhysRevLett.116.097204,PhysRevLett.120.207204}. Of a particular interest is the behavior of a magnonic spin current 
 flowing  from a ferromagnet (FM) across an insulating AMF layer \cite{PhysRevLett.113.097202,PhysRevB.93.054412,PhysRevLett.116.186601,doi:10.1063/1.4918990,Li2020,Vaidya160,Lebrun2018,PhysRevLett.124.217201}, which offers a way for  integration of magnonic spintronic and AFM devices, and allowing to act on the AFM by exciting the FM layer \cite{PhysRevB.93.054412}.

The present work proves that, due to finite temperature magnonic excitations in AFM, the  electronic spin-current in heavy metal (HM) can traverse through the insulating AFM layer.  In our model, not FM, but the AFM layer plays the role of the spin-current tunnel junction. We sandwich the AFM layer between two HM layers, meaning we consider HM / insulating AFM / HM heterostructure.  At finite temperatures, the electronic spin current creates (annihilates) thermal magnons with spin polarization opposite (parallel) to the polarization of the electron spin. Thus, spins of the electrons from the HM layer are efficiently converted to  AFM magnons. The induced magnon spins are further delivered to the second HM layer,  and eventually are  converted back to the electron spin current via the spin pumping effect. The effectiveness of the spin transport via magnons depends on the electronic spin current and on the external magnetic field.

The paper is organized as follows: In section \textbf{II} we specify the model,
in section \textbf{III} we analyze the  polarization of magnons and derive the magnon eigenmodes,
in section \textbf{IV} we explore the effect of STT. The thermal and spin pumping effects we discuss in section \textbf{V}. In sections  \textbf{VI} and \textbf{VII} we discuss the effect of external magnetic fields and conclude with section \textbf{VIII}.

\section{Theoretical model}
\begin{figure}[htbp]
	\includegraphics[width=0.48\textwidth]{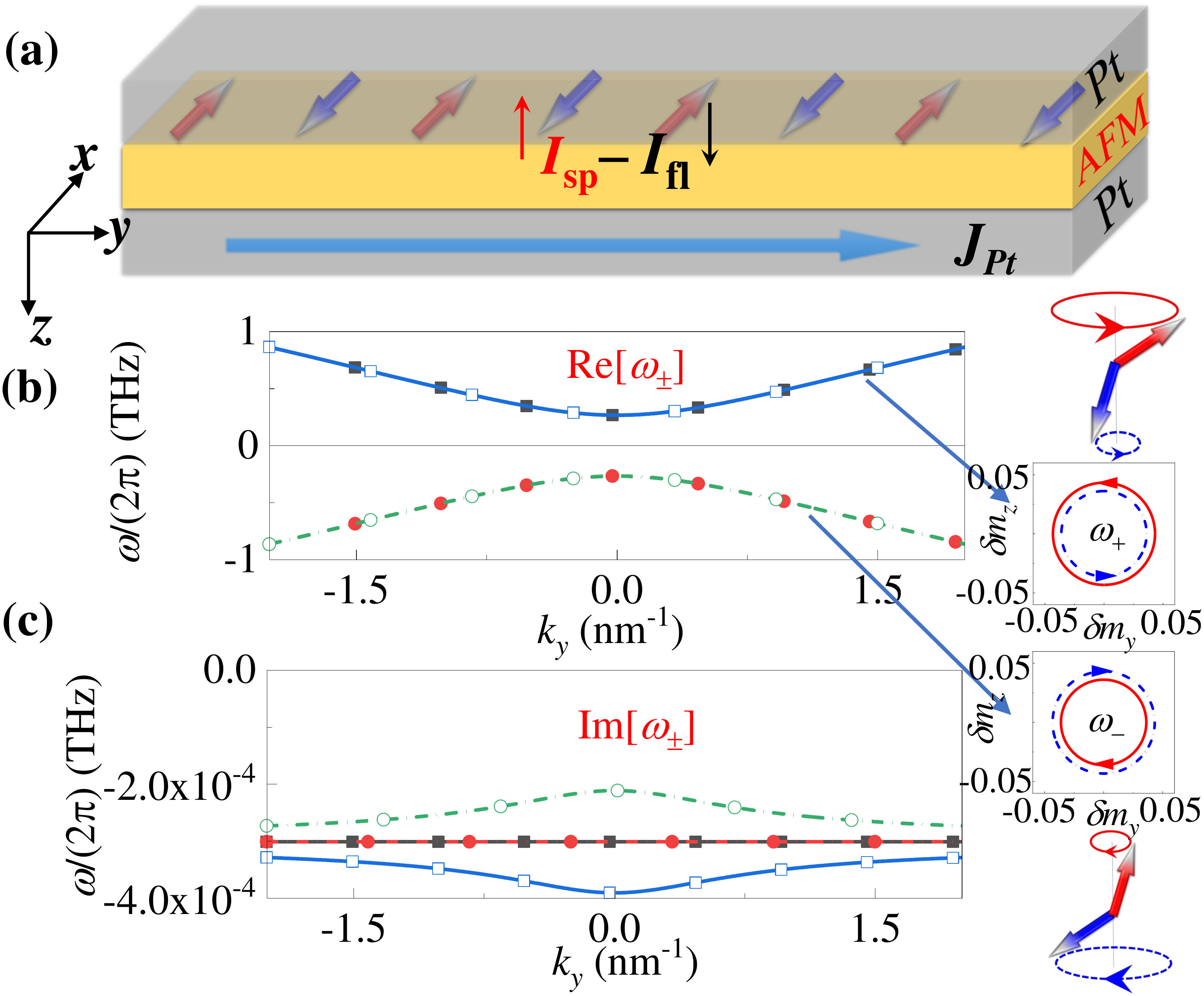}
	\caption{\label{model} (a) Schematics of the studied heterostructure. The insulating antiferromagnet is sandwiched between two HMs (here, Pt). An charge current with density $ J_{{\rm Pt}} $ flows in one of the metal layer.  The pure electron spin current generated through the spin Hall effect is denoted by its spin polarization direction $ \vec{\mu}_s^N = -\vec{z} \times \vec{j}_{{\rm Pt}} $. (b) Dispersion relations (i.e. real parts) and (c) imaginary parts of eigenfrequencies $ \omega_{+} $ (squares) and  $ \omega_{-} $ (circles), with ($ c_J = 0.0001 $ THz, open dots) and without STT ($ c_J = 0 $, solid dots). Bottom right panel shows the magnon precession trajectories in the $ y $-$ z $ plane for $ k_y = 0 $. Magnons are excited by a microwave field $\vec{\omega}_H = (0, h_y, h_z)e^{(i k_y y - i \omega t)} $ with $ h_y = 3 \times 10^{-4} $ THz, $ h_z = 3 \times 10^{-4} i $ THz,  and $ \omega  = {\rm  Re}[\omega_{\pm}] $. Red and blue arrows represent the sublattice magnetizations $ \vec{m}_1 $ and $ \vec{m}_2 $, respectively.}
\end{figure}

The structure considered in this study is shown in Fig. \ref{model}(a). The AFM layer is sandwiched between two HMs. A charge current with density $ J_{Pt} $ passes through one metal and induces a transversal spin current due to the spin Hall effect. The electronic spin current is converted into a magnonic spin current at the AFM/HM interface via the SHE-based spin-transfer torque (STT) \cite{PhysRevLett.120.207204, PhysRevLett.116.207603}. The magnonic spin current is detected in the second metal via the spin pumping effect \cite{PhysRevB.95.220408, Li2020, Vaidya160,chotorlishvili2015electromagnetically}.

To describe the magnetization dynamics in the AFM, we introduce the average magnetization vector $ \vec{m} = (\vec{m}_1 + \vec{m}_2)/2 $ and the N$ {\rm \acute{e}} $el  vector $ \vec{n} = (\vec{m}_1 - \vec{m}_2)/2 $. Here, $ \vec{m}_1 $ and $ \vec{m}_2 $ represent sublattice magnetizations of the AFM under the constraints $ |\vec{m}|^2 + |\vec{n}|^2 = 1 $ and $ \vec{m} \cdot \vec{n} = 0 $. The dynamics of $ \vec{m} $ and $ \vec{n} $ are governed by the stochastic Landau-Lifshitz-Gilbert (LLG) equations with STTs \cite{PhysRevB.95.220408}
 \begin{equation}
\begin{small}
\begin{aligned}
\displaystyle  \partial_t \vec{m} &= \frac{1}{2}(\vec{\omega}_m \times \vec{m} + \vec{\omega}_n\times \vec{n}) + \vec{\tau}_m^T + \vec{\tau}_m^{{\rm GD}} + \vec{\tau}_m^{{\rm STT}} ,\\
\partial_t \vec{n} &= \frac{1}{2}(\vec{\omega}_m \times \vec{n} + \vec{\omega}_n\times \vec{m}) + \vec{\tau}_n^T + \vec{\tau}_n^{{\rm GD}} + \vec{\tau}_n^{{\rm STT}}.
\label{llg}
\end{aligned}
\end{small}
\end{equation}
The frequencies $ \vec{\omega}_m $ and $ \vec{\omega}_n $ represent the effective fields and are defined through $ \vec{\omega}_m = -\frac{\gamma}{M_s} \frac{\delta E_{{\rm AFM}}}{\delta \vec{m}} $ and the $ \vec{\omega}_n = -\frac{\gamma}{M_s} \frac{\delta E_{{\rm AFM}}}{\delta \vec{n}} $. The free energy density \cite{PhysRevB.97.054423} $ E_{{\rm AFM}} $ has the form:
\begin{equation}
	\begin{small}
		\begin{aligned}
			\displaystyle  E_{{\rm AFM}} = &\frac{M_s}{\gamma}\{\omega_E (\vec{m}^2 - \vec{n}^2) -a^2\frac{\omega_E}{4}[(\vec{\nabla} \vec{m})^2 - (\vec{\nabla} \vec{n})^2] \\
			&- 2 \vec{\omega}_H \cdot \vec{m} - \omega_A(m_x^2 + n_x^2)\},
			\label{energy}
		\end{aligned}
	\end{small}
\end{equation}
where $ \omega_E $ is the exchange frequency, $ \omega_A $ is the easy-axis (along $ \textbf{x} $) anisotropy frequency, $ \vec{\omega}_H $ is the frequency describing the external magnetic field, and $ a $ is the length of the antiferromagnetic unit cell. In the stochastic LLG equations (\ref{llg}), the temperature is introduced by the thermal random magnetic field torque 
 \begin{equation}
\begin{small}
\begin{aligned}
\displaystyle  \vec{\tau}_m^T & = \vec{h}_m \times \vec{m} + \vec{h}_n \times \vec{n},\\
\vec{\tau}_n^T & = \vec{h}_m \times \vec{n} + \vec{h}_n \times \vec{m}.
\label{heat}
\end{aligned}
\end{small}
\end{equation}
The Gilbert damping torques are given through
 \begin{equation}
\begin{small}
\begin{aligned}
\displaystyle  \vec{\tau}_m^{{\rm GD}} & = \alpha (\vec{m} \times \partial_t \vec{m} + \vec{n} \times \partial_t \vec{n}),\\
\vec{\tau}_n^{{\rm GD}} & = \alpha (\vec{m} \times \partial_t \vec{n} + \vec{n} \times \partial_t \vec{m}),
\label{damping}
\end{aligned}
\end{small}
\end{equation}
and the STTs read:
 \begin{equation}
\begin{small}
\begin{aligned}
\displaystyle  \vec{\tau}_m^{{\rm STT}} & = c_J(\vec{m} \times \vec{\mu}_s^N \times \vec{m}  + \vec{n} \times \vec{\mu}_s^N \times \vec{n} ),\\
\vec{\tau}_n^{{\rm STT}} & = c_J(\vec{n} \times \vec{\mu}_s^N \times \vec{m} + \vec{m} \times \vec{\mu}_s^N \times \vec{n}).
\label{stt}
\end{aligned}
\end{small}
\end{equation}
Here $ \alpha $ is the Gilbert damping constant, the strength of STT is quantified through  $ c_J = \frac{2 \gamma \hbar \theta_{{\rm SH}} \lambda G_r \tanh(d_{{\rm Pt}}/2\lambda) J_{{\rm Pt}}}{e \mu_0 d_{{\rm AF}}M_s[\sigma + 2 \lambda G_r \coth(d_{{\rm Pt}}/\lambda)]} $ \cite{PhysRevB.87.144411,PhysRevB.97.094401}, thermal fields $ \vec{h}_m $ and  $ \vec{h}_n $ satisfy the time correlation \cite{PhysRevB.81.214418} $ \langle h_{m,p}(\vec{r},t) h_{m,q}(\vec{r}',t') \rangle = \langle h_{n,p}(\vec{r},t) h_{n,q}(\vec{r}',t') \rangle = \delta_{pq} \delta(t - t') \delta(\vec{r} - \vec{r}') \frac{\alpha \gamma k_B T}{\mu_0 M_s V} $. $ \vec{\mu}_s^N = (1,0,0) $ represents the electron spin polarization, $ J_{{\rm Pt}} $ is the electric current density, $ \theta_{{\rm SH}} $ is the spin Hall angle, $ \sigma $ is the electric conductivity, $ \lambda $ is the spin diffusion length, $ G_r $ is the spin mixing interface conductance per unit area, $ d_{{\rm Pt}} $ and $ d_{{\rm AF}} $ are the thicknesses of Pt and AFM, $ k_B $ is the Boltzmann constant, $ T $ is the temperature, $ V $ is the volume of AFM, $ \gamma $ is the gyromagnetic ratio, $ M_s $ is the saturation magnetization of sublattice, and $ p,q = x,y,z $.

\section{Magnon polarization}
To construct an analytic model for describing the propagation of magnons in the AFM we consider slight derivations from the stable state ($ \vec{m}_0 = (0, 0, 0) $ and $ \vec{n}_0 = (1, 0, 0) $), $ \vec{m} = \vec{m}_0 + (0, \delta m_y, \delta m_z) $, and $ \vec{n} = \vec{n}_0 + (0, \delta n_y, \delta n_z) $. The eigen solutions of the linearized equation (\ref{llg}) have the form: $ \delta m_p = X_p e^{i(k_y y - \omega t)} $ and $ \delta n_p = Y_p e^{i(k_y y - \omega t)} $, where $ p = y, z $. We insert $ \vec{m} $ and $ \vec{n} $ into equation (\ref{llg}) and obtain the equation $ i \partial_t \vec{\psi} = \hat{H} \vec{\psi} $ for the vector $ \vec{\psi} = (X_y, X_z, Y_y, Y_z) $. 
With the definitions of the frequencies $ \omega_k = 2 \omega_E + \omega_A - \frac{a^2 \omega_E k_y^2}{4} $ and $ \omega_{ak} = \omega_A + \frac{a^2 \omega_E k_y^2}{4} $, 
the Hamiltonian $ \hat{H} $ (without external excitation) reads
\begin{equation}
\begin{small}
\begin{aligned}
\displaystyle \hat{H} = \left( \begin{matrix} -i \alpha \omega_k  & i \alpha c_J  & -i c_J & -i \omega_{ak} \\ -i \alpha c_J & -i \alpha \omega_k & i \omega_{ak} & -i c_J \\-i c_J & -i \omega_k & -i \alpha \omega_{ak} & i \alpha c_J \\ i \omega_k & -i c_J & -i \alpha c_J & -i \alpha \omega_{ak} \end{matrix} \right).
\label{Ham}
\end{aligned}
\end{small}
\end{equation}
From this Hamiltonian two eigen frequencies $ \omega_{\pm} $ follows
\begin{equation}
\begin{small}
\begin{aligned}
\displaystyle  \omega_{\pm} =& \pm\sqrt{(\omega_{ak} - i c_J)( \omega_k - i c_J) - \alpha^2 \left( \omega_E - \frac{a^2\omega_E k_y^2}{4}\right) ^2} \\
 &- \alpha c_J - i \alpha (\omega_{ak} + \omega_k)/2.
\label{eigen0}
\end{aligned}
\end{small}
\end{equation}
The spin-wave modes $ \omega_{\pm} $ correspond to the opposite circular polarizations. Without STT ($ c_J = 0 $), the two magnon modes are degenerate, and the magnon dispersion relations $ {\rm  Re}[\omega_{\pm}] = \pm\sqrt{\omega_{ak} \omega_k - \alpha^2 (\omega_E - \frac{a^2\omega_E k_y^2}{4})^2} $ of the two modes are symmetric with respect to $ \omega = 0 $. Using $ {\rm MnF}_2 $  as a material with the parameters  $ \omega_E = 9.3 \times 10^{12} {\rm s}^{-1} $, $ \omega_A = 1.5 \times 10^{11} {\rm s}^{-1} $, $ M_s = 48000 $ A/m, $ \alpha = 0.0002 $, and $ a = 4.1 {\rm \AA} $, we  calculated numerically  the degenerate modes $ \omega_{\pm} $ for $ c_J = 0 $. The results are shown in Fig. \ref{model}(b-c).

To explore the polarization of the two magnon modes, we apply the microwave filed $\vec{\omega}_H = (0, h_y, h_z)e^{(i k_y y - i \omega t)} $. The excited magnon amplitudes are extracted analytically from the linearized equation (\ref{llg}) as 
\begin{widetext}
\begin{equation}
\begin{small}
\begin{aligned}
\displaystyle   X_y &= \frac{(h_y + ih_z )(-ic_J - \omega_{ak} + i \alpha \omega)}{l^\omega_+} - \frac{(h_y - ih_z )(-ic_J + \omega_{ak} - i \alpha \omega)}{l^\omega_-},\\
X_z &= \frac{(-i h_y + h_z )(-ic_J - \omega_{ak} + i \alpha \omega)}{l^\omega_+} - \frac{(ih_y + h_z )(-ic_J + \omega_{ak} - i \alpha \omega)}{ l^\omega_-},\\
Y_y &= \frac{(h_y + ih_z ) \omega}{l^\omega_+} - \frac{(h_y - ih_z ) \omega}{l^\omega_-},\\
Y_z &= \frac{(-ih_y + h_z ) \omega}{l^\omega_+} - \frac{(ih_y + h_z ) \omega}{l^\omega_-},
\label{excite}
\end{aligned}
\end{small}
\end{equation}
\end{widetext}
with $ l^\omega_{\pm} := 2\omega^2  + 2[-c_J \pm i(\omega_{ak} - i \alpha \omega) ][-c_J \pm i (\omega_k - i \alpha \omega)]$. Excited by the microwave field, the local magnetization precesses around the equilibrium state. The precessions of the two sublattices are demonstrated in the bottom right panel in Fig. \ref{model} for the case $ k_y = 0 $.  The magnons of the two modes $ \omega_{\pm} $ have opposite chirality. For the mode $ \omega_+ $, the magnons of both $ \vec{m}_1 $ and $ \vec{m}_2 $ sublattices,  precess around $ +\textbf{x} $ direction counterclockwise, meaning that the right-hand polarized magnon (identified with $ \vec{m}_1 $) is coupled to left-hand magnon (identified with $ \vec{m}_2 $), and the amplitude of magnons for $ \vec{m}_1 $ is larger. For the $ \omega_- $ mode, the circular polarizations of both magnons are reversed (clockwise precession around $ +x $), and the amplitude of the  right-hand magnons for $ \vec{m}_2 $ is larger.

\section{Effect of STT}
\begin{figure}[htbp]
	\includegraphics[width=0.48\textwidth]{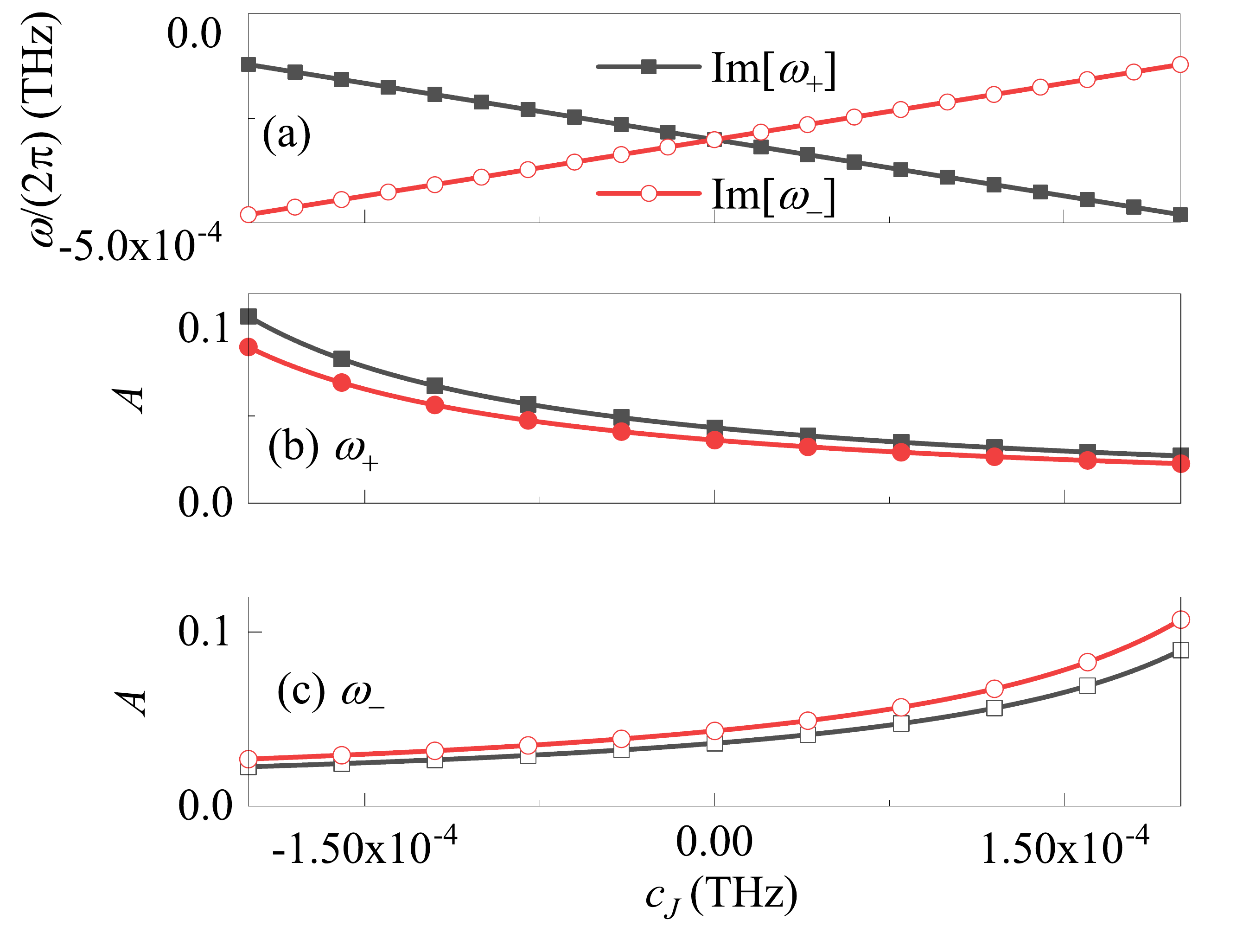}
	\caption{\label{ce-amp-f} (a) $ c_J $ dependence of the imaginary parts of eigen frequencies $ \omega_{\pm} $ at $ k_y = 0 $.  (b-c) Excited by $\vec{\omega}_H = (0, h_y, h_z)e^{(i k_y y - i \omega t)} $ with $ h_y = 3 \times 10^{-4} $ THz, $ h_z = 3i \times 10^{-4} $ THz and $ k_y = 0 $, $ c_J $ dependence of excited oscillation amplitude for (b) $ \omega  = {\rm  Re}[\omega_+] $ and (c) $ \omega  = {\rm  Re}[\omega_-] $. Solid squares and open circles are respectively the amplitudes of the sublattice magnetization $ \vec{m}_1 $ and $ \vec{m}_2 $. }
\end{figure}

 The influence of the SHE-induced STT on the eigenfrequencies $ \omega_{\pm} $ is shown in Figs. \ref{model}(b-c).  The changes due to STT in the real parts of $ \omega_{\pm} $  are negligible. STT with $ c_J > 0 $ (i.e. electron polarization  points along the $ +x $ direction), increases  $ {\rm  Im}[\omega_+] $, meaning that the attenuation of this mode is enhanced.  On the other hand $ {\rm  Im}[\omega_-] $ decreases, and the attenuation is weakened. The change in $ {\rm  Im}[\omega] $ depends linearly on $ c_J $ (Fig. \ref{ce-amp-f}(a)), and $ {\rm  Im}[\omega_+] $ ($ {\rm  Im}[\omega_-] $), is decreased (increased) by reversing the STT ($ c_J < 0 $).

The STT also affects the magnon excitation efficiency. Using polarized microwave field $\vec{\omega}_H = (0, h_y, h_z)e^{(i k_y y - i \omega t)} $ ($ h_y = 3 \times 10^{-4} $ THz, $ h_z = 3i \times 10^{-4} $ THz, $ k_y = 0 $, and $ \omega  = {\rm  Re}[\omega_\pm] $) and Eq. (\ref{excite}), we calculate  amplitudes of the excited magnetization oscillations, see Fig. \ref{ce-amp-f}. Apparently the positive (negative) $ c_J $ decreases (increases) the efficiency of exciting magnons of $ \omega_+ $ mode. The effect is reversed for $ \omega_- $ mode. In view of the changes in the imaginary parts of eigenfrequencies (Fig. \ref{ce-amp-f}(a)), we conclude that the effectiveness of exciting magnon increases if the imaginary parts of $ \omega_{\pm} $ is decreased, i.e., the effective magnon damping $ \alpha_{{\rm eff}} $ is lowered.  

Enhancement and annihilation of magnons due to the STT are asymmetric and nonlinear, and the increase of the magnon amplitude is much stronger.  This finding is in line with the STT induced magnon enhancement/annihilation in FM, where the enhancement (annihilation) occurs when electron polarization is opposite (parallel) to the magnon polarization and decreases (increases) the magnon effective damping \cite{PhysRevB.95.020414, PhysRevB.97.094401}.

\section{Thermal effect and spin pumping}
  From the Eq. (\ref{llg}) we infer that the thermal magnons are excited by the thermal random fields $ \vec{h}_m $ and  $ \vec{h}_n $.  By solving the  dynamic equations, we obtain the dynamic magnetic susceptibility matrix $ \hat{\chi}(\omega)$ , and $ l_p(\omega) = \sum_q \chi_{pq} h_q(\omega) $ with $ l_1 = m_y $, $ l_2 = m_z $, $ l_3 = n_y $, $ l_4 = n_z $, $ h_1 = h_{m,y} $, $ h_2 = h_{m,z} $, $ h_3 = h_{n,y} $ and $ h_4 = h_{n,z} $):
\begin{equation}
\begin{small}
\begin{aligned}
\displaystyle \hat{\chi} = \left( \begin{matrix}  \frac{c_-}{l_+}-\frac{c_+}{l_-} & \frac{ic_-}{l_+}+\frac{ic_+}{l_-}  & \frac{\omega}{l_+}-\frac{\omega}{l_-} & \frac{i\omega}{l_+}+\frac{i\omega}{l_-} \\ -\frac{ic_-}{l_+}-\frac{ic_+}{l_-} & \frac{c_-}{l_+}-\frac{c_+}{l_-} & -\frac{i\omega}{l_+}-\frac{i\omega}{l_-} & \frac{\omega}{l_+}-\frac{\omega}{l_-} \\ \frac{\omega}{l_+} - \frac{\omega}{l_-} & \frac{i \omega}{l_+} + \frac{i \omega}{l_-} & \frac{d_-}{l_+}-\frac{d_+}{l_-} & \frac{i d_-}{l_+}+\frac{id_+}{l_-} \\ -\frac{i \omega}{l_+} - \frac{i \omega}{l_-} & \frac{\omega}{l_+} - \frac{\omega}{l_-} & -\frac{i d_-}{l_+}-\frac{id_+}{l_-} & \frac{d_-}{l_+}-\frac{d_+}{l_-} \end{matrix} \right),
\label{chi}
\end{aligned}
\end{small}
\end{equation}
where  $ c_{\pm} := -i c_J \pm (\omega_{ak} - i \alpha \omega) $, and $ d_{\pm} := -i c_J \pm (\omega_k - i \alpha \omega) $.

\begin{figure}[htbp]
	\includegraphics[width=0.48\textwidth]{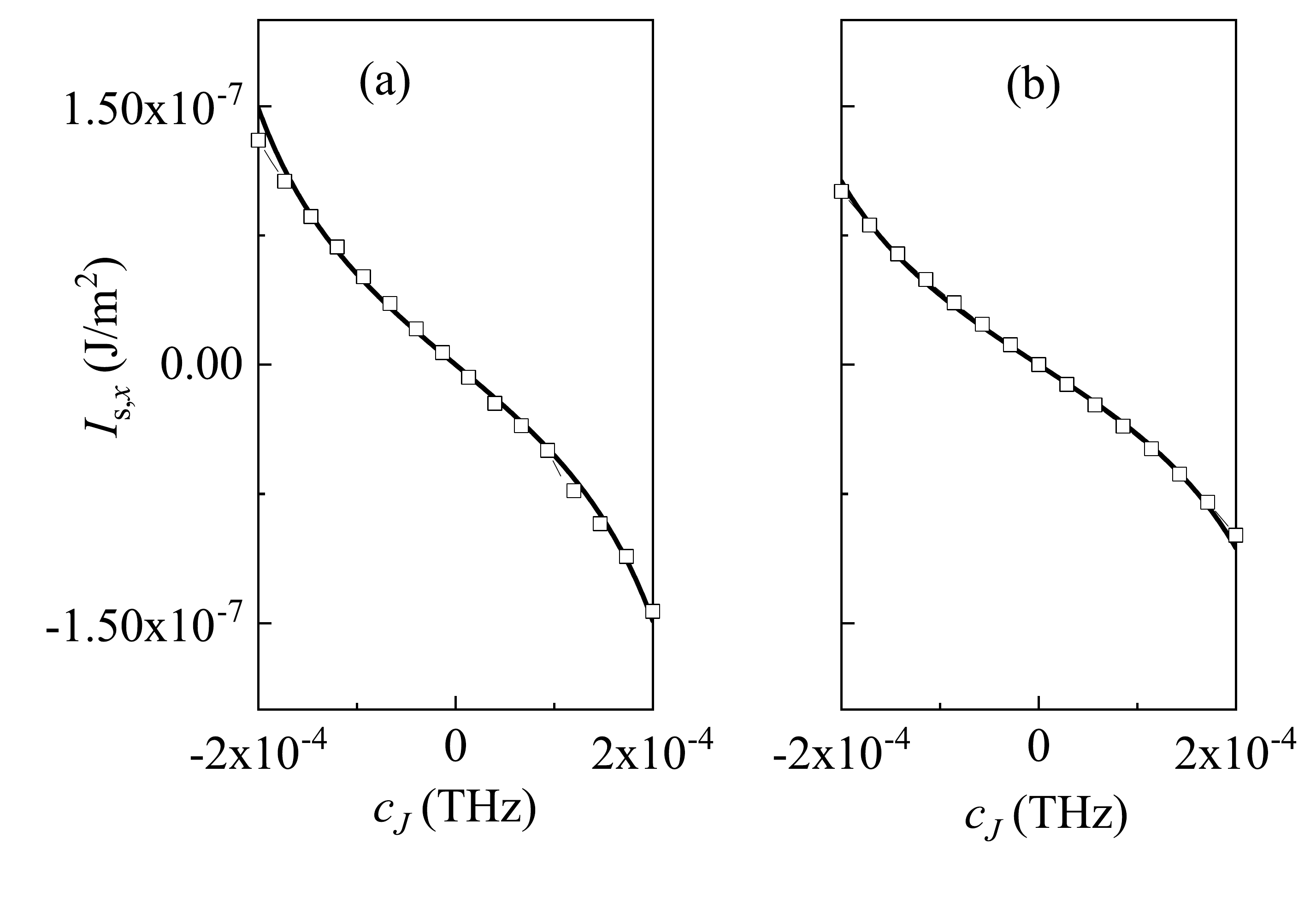}
	\caption{\label{cj-pumpf-h0}  $ c_J $ dependence of net current $ \langle I_{s,x} \rangle $ at   $ T = 30 $ K
		for (a) macrospin and (b) 1D models.  Analytical (solid lines) and numerically simulated (open dots) results are shown. }
\end{figure}

By virtue of spin pumping,  the magnetization dynamics in the AFM can pump into the neighboring metal layer the spin current\cite{PhysRevB.95.220408} 
\begin{equation}
\begin{small}
\begin{aligned}
\displaystyle  \vec{I}_{{\rm sp}} = \frac{\hbar g_r}{2 \pi} (\vec{m} \times \partial_t \vec{m} + \vec{n} \times \partial_t \vec{n}).
\label{Isp}
\end{aligned}
\end{small}
\end{equation}
Here, $ g_r = G_r h /e^2 $ is the rescaled interface mixing conductance. 
In contrast  to  spin pumping current, the fluctuation spin current $ \vec{I}_{{\rm fl}} = -\frac{2M_s}{\gamma} (\vec{m} \times \vec{h}'_m + \vec{n} \times \vec{h}'_n) $ flows back to the AFM \cite{PhysRevB.81.214418}. $ \vec{h}'_m $ and $ \vec{h}'_n $ satisfy the time correlation $ \langle h'_{m,p}(\vec{r},t) h'_{m,q}(\vec{r}',t') \rangle = \langle h'_{n,p}(\vec{r},t) h'_{n,q}(\vec{r}',t') \rangle = \delta_{pq} \delta(t - t') \delta(\vec{r} - \vec{r}') \frac{\alpha' \gamma k_B T}{\mu_0 M_s V} $, and $ \alpha'=\gamma \hbar g_r /(4 \pi M_s V) $. The net spin current injected into the neighboring metal is $ \vec{I}_s = \vec{I}_{{\rm sp}} + \vec{I}_{{\rm fl}} $ . Then, with the magnon dynamics described by Eq. (\ref{chi}), the time derivative of the correlation function for $ \vec{I}_{{\rm sp}} $ within the macrospin model ($ k_y = 0 $) is  
\begin{equation}
\begin{small}
\begin{aligned}
\displaystyle  \langle \dot{l}_p l_q \rangle = \sigma^2_T \int i \omega \sum_n \chi_{pn}(\omega) \chi_{qn}(-\omega) \frac{d\omega}{2\pi}.
\label{correIsp}
\end{aligned}
\end{small}
\end{equation}
 $ p,q,n =1,2,3,4 $. Using the contour integration method, we derive the finite $ x $ component of the spin current $ I_{{\rm sp,x}} $ as 
\begin{equation}
\begin{small}
\begin{aligned}
\displaystyle  \langle I_{{\rm sp,}x} \rangle = \frac{\hbar g_r \sigma^2_T c_J (\omega_{ak} + \omega_k)}{2 \pi (c_J^2 - \alpha^2 \omega_{ak} \omega_k)}.
\label{aIsp}
\end{aligned}
\end{small}
\end{equation}
The two other components of the spin current vanish $ \langle I_{sp,y} \rangle = \langle I_{sp,z} \rangle = 0 $. 
The above equation indicates  that the spin pumping current's polarization is parallel to the electron polarization and  the equilibrium N$ {\rm \acute{e}} $el order vector.  Without STT, two degenerate magnon modes are equally excited by thermal fluctuation, and the pumping current $ \langle I_{sp,x} \rangle = 0 $.  Below the critical value $ c_J < \alpha \sqrt{\omega_A \omega_k} $ (above which the STT changes the stable state), positive $ c_J $ creates the negative $ \langle I_{sp,x} \rangle $. The reason is that positive $ c_J $ enhances the $ \omega_- $ mode polarized towards $ -x $ (generating negative pumping current) but weakens the $ \omega_+ $ mode polarized towards $ +x $ (generating positive pumping current). The sign of $ \langle I_{{\rm sp,}x} \rangle $ changes with reversing the direction of the electric current (the sign of $ c_J $). The change of the magnon density induced by STT is nonlinear (see Fig. \ref{ce-amp-f}), and therefore the amplitude of $ \langle I_{sp,x} \rangle $ also changes nonlinearly with $ c_J $.  The fluctuating spin current vanishes $  \vec{I}_{{\rm fl}} = 0 $. Thus, the net current is equal to the spin pumping current $ \langle I_{s,x} \rangle = \langle I_{sp,x} \rangle $. Numerical results calculated from Eq. (\ref{aIsp}) are shown in Fig. \ref{cj-pumpf-h0}(a) and support the analytical findings.

We extended the results obtained for the single macrospin model  to the  spatially inhomogeneous dynamical modes (i.e., excited finite wavevector $ \vec{k} $), and obtain the time derivative of the correlation function $ \vec{I}_{{\rm sp}} $
\begin{equation}
\begin{small}
\begin{aligned}
\displaystyle  \langle \dot{l}_p l_q \rangle = \sigma^2_T S_\zeta \int\frac{d^{\zeta} \vec{k}}{(2 \pi)^{\zeta}} \int i \omega \sum_n \chi_{pn}(\omega,\vec{k}) \chi_{qn}(-\omega,-\vec{k}) \frac{d\omega}{2\pi}.
\label{correIsp1d}
\end{aligned}
\end{small}
\end{equation}
Here, $ \zeta = 1 $ is for the one-dimensional (1D) model ($ \vec{k} = k_y \vec{e}_y $ and  $ S_1 $ is the sample length along $ y $). For the two-dimensional (2D) model $ \zeta = 2 $,  ($ \vec{k} = k_x \vec{e}_x + k_y \vec{e}_y $ and $ S_2 = S_{xy} $ is the area of the sample plane).

Due to the limited size of discrete unit cell, the value of magnon wavevector can not be too large. We integrate  Eq. (\ref{correIsp1d}) in the finite range of $ -k_c < k_{x,y} < k_c $ with $ k_c = \pi / l_y $, where $ l_y $ is the unit cell size. For a  1D model, we infer
\begin{equation}
\begin{small}
\begin{aligned}
\displaystyle  &\langle I_{{\rm sp,}x} \rangle = \hbar g_r \sigma_T^2 l_y  c_J (\omega_A + \omega_E) \times\\
 &\frac{( \omega_{j+} {\rm atan}( \frac{\alpha a k_c \sqrt{\omega_E}}{2\omega_{j-}}) -\omega_{j-}{\rm  atan}(\frac{\alpha a k_c \sqrt{\omega_E}}{2\omega_{j+}})))}{2\pi^2 \omega_{j+} \omega_{j-} \sqrt{\alpha^2(\omega_A + \omega_E)^2 - c_J^2}  },
\label{aIsp1d}
\end{aligned}
\end{small}
\end{equation}
with
\begin{eqnarray}\label{Here we define}
\omega_{j \pm} = \sqrt{\alpha (\pm \sqrt{\alpha^2(\omega_A + \omega_E)^2 - c_J^2} - \alpha \omega_E)}.
\end{eqnarray}
If STT is applied $ c_J\neq 0$, the variation of $ \langle I_{{\rm sp,}x} \rangle $ for 1D model is similar to that for macrospin, including the aspects of sign and nonlinearity, as confirmed by numerical calculations in Fig. \ref{cj-pumpf-h0}(b). 

 To quantify the effectiveness of the above conversion process, we compare the difference between the injected electronic spin current and output magnonic pumping current. Here, via $ J_s^x = -\frac{2 c_J}{\mu_0 d_{AF} M_s}  $, the $ x $ component of electronic spin current density  $ J_s^x $ is directly determined from the amplitude $ c_J $ of STT \cite{PhysRevB.87.144411}. With the above parameters, we find $ J_s^x = -5.5 \times 10^{-7}  $ J/m$ ^2 $ under $ c_J = 2 \times 10^{-4} $ THz; and in this case around 27 \% electronic spin current is converted into the magnonic pumping current in Fig. \ref{cj-pumpf-h0}.

To support the analytical results, we numerically solved for the LLG  Eqs. (\ref{llg}). In the numerical simulation the spin pumping current's value is determined from the expression Eq. (\ref{Isp}). Under the same parameters adopted above, we compare the simulation results with the analytical calculations in Fig. \ref{cj-pumpf-h0}, confirming the value of the analytical expressions. 
In addition, we consider the case with  the electron spin polarization $ \vec{\mu}_s^N = (0,1,0) $ being  perpendicular to the equilibrium N$ {\rm \acute{e}} $el order vector $ \vec{n}_0 $. Our calculations show that the STT does not affect the magnons in the AFM in this case, and therefore $ \langle I_{{\rm sp,}x} \rangle = 0 $. This conclusion is confirmed also by the numerical simulation.

\section{Applying magnetic field along \textit{x}}

\begin{figure}[htbp]
	\includegraphics[width=0.48\textwidth]{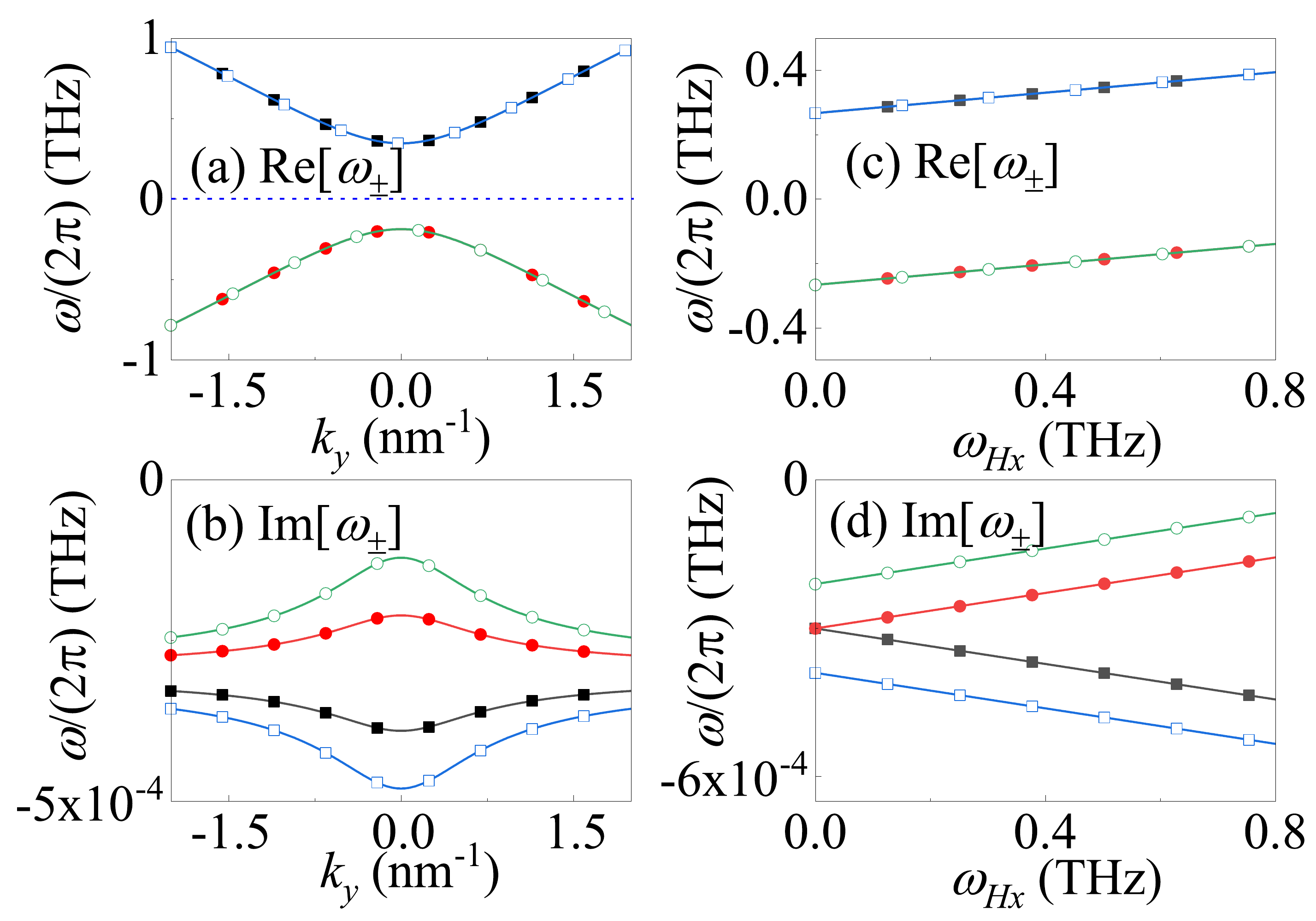}
	\caption{\label{w-hx-f}  (a) Dispersion relations (i.e. real parts) and (b) imaginary parts of eigenfrequencies of $ \omega_{+} $ (squares) and  $ \omega_{-} $ (circles) when $ \omega_{Hx} = 0.1$ THz and $ c_J = 0$ (solid dots) and $ c_J = 0.0001$ THz (open dots). (c) Real and (d) imaginary parts of eigenfrequencies $ \omega_{\pm} $ as functions of $ \omega_{Hx} $ under $ c_J = 0 $ (solid dots) and $ c_J = 0.0001$ THz (open dots).}
\end{figure}

An applied magnetic field impacts the magnon polarization and leads to the nontrivial phenomena of electron-magnon spin conversion in the AFM /heavy heterostructure. Here, we mainly consider the case with the external magnetic field $ \omega_{Hx} $ applied along the easy-axis ($ \textbf{x} $ axis), for $ \omega_{Hx} < \sqrt{2 \omega_E \omega_A} = 1.7 $ THz. In this range, the linear antiparallel structure ($ \vec{m}_0 = (0, 0, 0) $ and $ \vec{n}_0 = (1, 0, 0) $) is stable. In the same manner we obtain the eigenfrequencies $ \omega_{\pm} $:
\begin{equation}
\begin{small}
\begin{aligned}
\displaystyle  \omega_{\pm} &= \pm \sqrt{F_{\pm}} + \omega_{ch} - \frac{i \alpha}{2} \omega_{ck}^+,
\label{eigenhx}
\end{aligned}
\end{small}
\end{equation}
where
\begin{equation}
\begin{small}
\begin{aligned}
 F_{\pm} = &\frac{1}{2\omega_{ch}}(-4 c_J^2 \omega_{Hx} + \alpha^3 c_J [4 \omega_{Hx}^2 + (\omega_{ck}^{-})^2] \\
& + \alpha^2 \omega_{Hx} [8 c_J^2 - 4 \omega^2 - (\omega_{ck}^{-})^2] + 4 \omega_{Hx} \omega_k \omega_A \\
& - 4 i c_j \omega_{ch} \omega_{ck}^{+} +4 \alpha [c_J^3 - i\omega_{Hx} \omega_{ch} \omega_{ck}^{+}\\
& -c_J (2 \omega_{Hx}^2 + \omega_k \omega_{ak} )] ),
\label{eigenhx0}
\end{aligned}
\end{small}
\end{equation}
 where $ \omega_{ch} = \omega_{Hx} - \alpha c_J $, and $ \omega_{ck}^{\pm} = \omega_k \pm \omega_{ak} $. For the parameters considered above, we calculate the magnon dispersion relations (Fig. \ref{w-hx-f}(a-b)). The real parts of both modes $ \omega_+ $ and $ \omega_- $  are shifted upward and the values of the corresponding imaginary parts are changed,  steering the separation between $ {\rm  Im}[\omega_{\pm}] $. These changes increase linearly   with $ \omega_{Hx}$, as demonstrated in Fig. \ref{w-hx-f}(c-d). STT  affects mainly the imaginary parts, and $ {\rm  Im}[\omega_+] $ ($ {\rm  Im}[\omega_-] $) is increased (decreased) by a positive $ c_J $ (cf. the calculated results in Fig. \ref{w-hx-f}).

\begin{figure}[htbp]
	\includegraphics[width=0.48\textwidth]{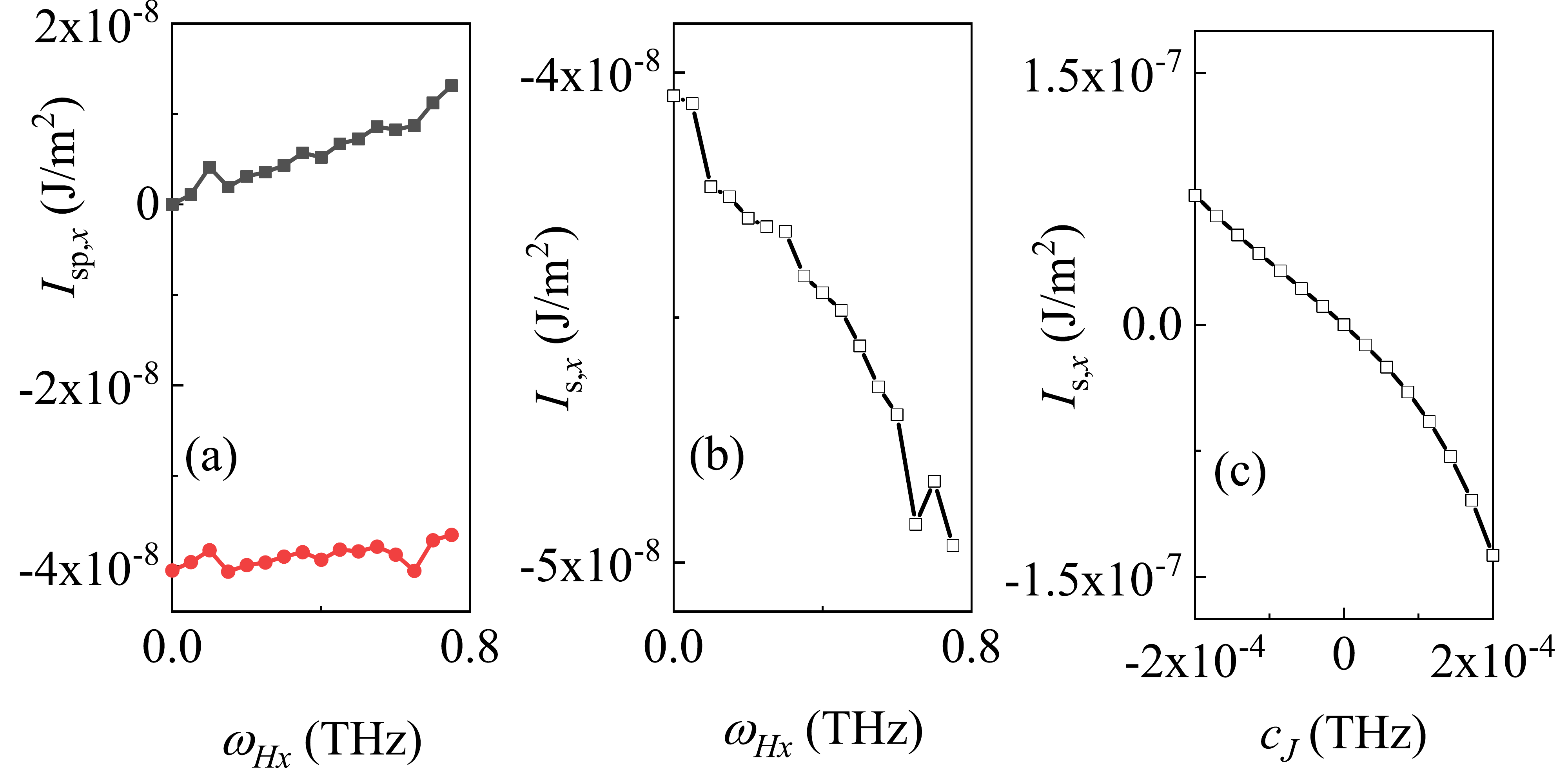}
	\caption{\label{pump-hx}  For a 1D model at the temperature $ T = 30 $ K, (a) spin pumping spin current $\langle I_{sp,x} \rangle$ (black squares for $ c_J = 0 $, red circles for $ c_J = 1 \times 10^{-4} $ THz) and (b) net spin current $\langle I_{s,x} \rangle$  (when $ c_J = 1 \times 10^{-4} $ THz) as functions of magnetic field $ \omega_{Hx} $. (c) At $ \omega_{Hx} = 0.7 $ THz, $ c_J $ dependence of net current  $\langle I_{s,x} \rangle$.  }
\end{figure}

The magnetic field induces a separation between the two degenerate modes $ \omega_{\pm} $, and at a finite temperature leads to a nonzero pumping current $ I_{sp,x} $ along the external magnetic field. The fluctuation spin current $ \langle I_{fl,x} \rangle $ is opposite to the pumping current $ \langle I_{sp,x} \rangle $, and the net current $ \vec{I}_s $ is 0 if  STT is not applied ($ c_J = 0 $).  STT can further enhance one of two thermal magnon modes and weaken the other one, generating a nonzero $ \vec{I}_s $, see Fig. \ref{pump-hx}(b). Surprisingly, the negative current $\langle I_{s,x}\rangle$  induced by the positive $ c_J $ also increases with $ \omega_{Hx} $. When calculating the dependence of the net current  $\langle I_{s,x} \rangle$  on the $ c_J $ at finite $ \omega_{Hx} = 0.7 $ THz, we find that the positive $ \omega_{Hx} $ enhances the negative  $\langle I_{s,x} \rangle$, while it weakens the positive  $\langle I_{s,x} \rangle$, as compared to the case $ \omega_{Hx} = 0 $ (Fig. \ref{cj-pumpf-h0}(b)). This effect leads to an asymmetric spin pumping in AFM  induced by STT, and acts in favor of converting the electronic current into a magnonic spin current and vice versa.
 

 \section{External magnetic field along \textit{y}}

Under the influence of a strong magnetic field $ \omega_{Hx} $ applied along the $ \textbf{x} $ axis or $ \omega_{Hy} $ applied along $ \textbf{y} $ axis, the linear antiparallel orientation of AFM magnetization  loses its stability and a spin-flop transition occurs. In this case, a nonzero net magnetization builds up along the magnetic field and increases with the amplitude of the magnetic field. The equilibrium N$ {\rm \acute{e}} $el order vector is perpendicular to the magnetic field. To explore the influences of the spin-flop on the net magnetization, we mainly study the case when $ \omega_{Hy} $ is applied along the $ \textbf{y} $ axis.

\begin{figure}[htbp]
	\includegraphics[width=0.48\textwidth]{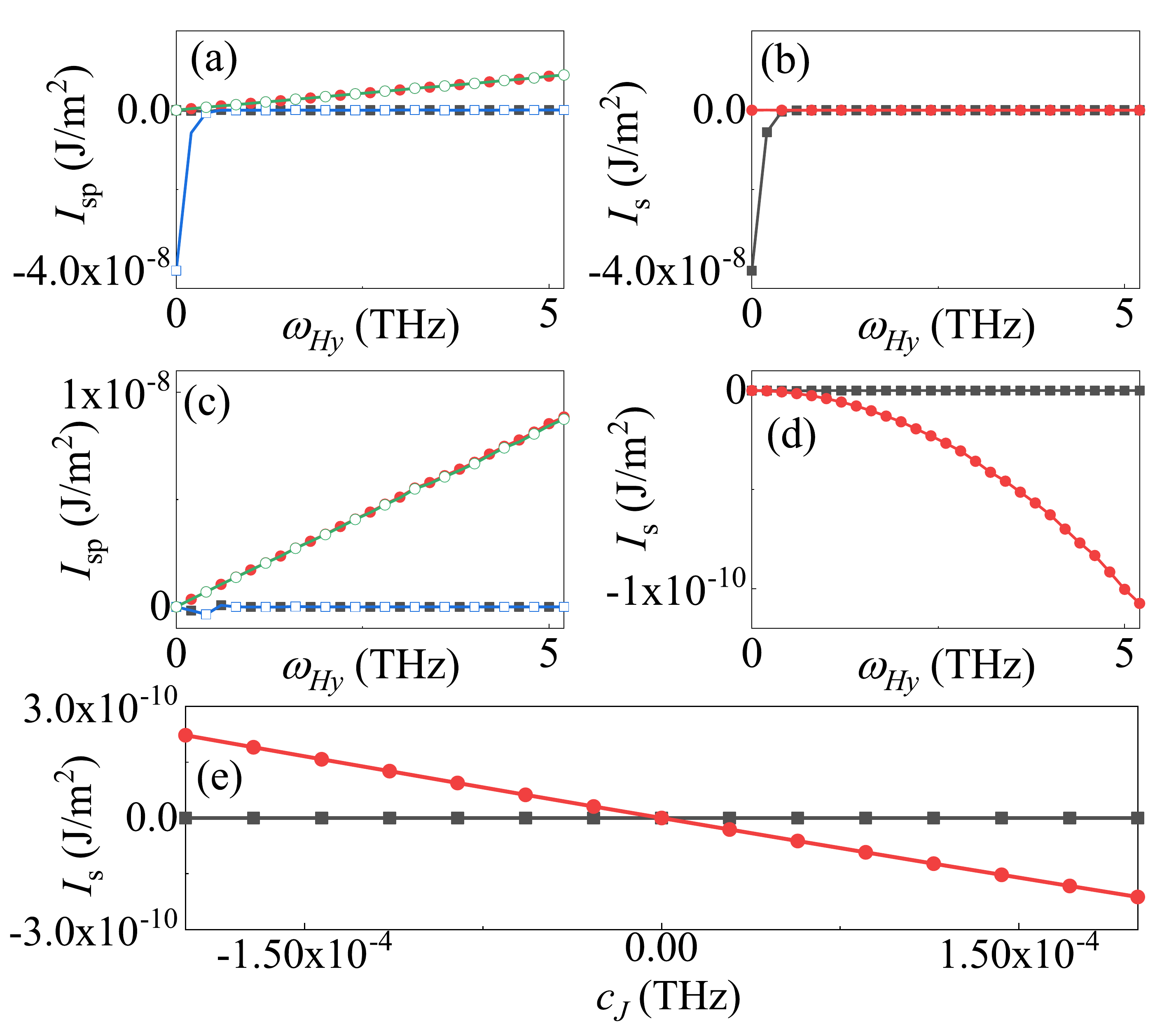}
	\caption{\label{pump-hy}  For 1D model at  $ T = 30 $ K, the $ x $ component $\langle I_{sp,x} \rangle$ (squares line) and the $ y $ component $\langle I_{sp,y} \rangle$ (circles line) of the spin pumping currents are shown as functions of the magnetic field $ \omega_{Hy} $, under  $ c_J = 0 $ (solid dots) or $ c_J = 1 \times 10^{-4} $ THz (open dots) with (a) $ \vec{\mu}_s^N = \vec{x} $ and (c) $ \vec{\mu}_s^N = \vec{y} $. With $ c_J = 1 \times 10^{-4} $ THz and (b) $ \vec{\mu}_s^N = \vec{x} $ and (d) $ \vec{\mu}_s^N = \vec{y} $, the net spin currents $\langle I_{s,x} \rangle$ (squares line) and $\langle I_{s,y} \rangle$ (circles line) as functions of the magnetic field $ \omega_{Hy} $.  (e) When $ \omega_{Hy} = 5.2  $ THz, $ c_J $ dependence of the net spin currents $\langle I_{s,x} \rangle$ (squares line) and $\langle I_{s,y} \rangle$ (circles line). }
\end{figure}

Applying positive $ \omega_{Hy} $ generates a net magnetization $ \vec{m}_0 = (0, m_{0,y},0) $ with $ m_{0,y} > 0 $. The thermal fluctuation of this net magnetization exerts a positive pumping current $\langle I_{sp,y} \rangle$ along $ y $, as shown by Fig. \ref{pump-hy}. 
This behavior is similar to the features of a thermal pumping current in FM \cite{PhysRevB.81.214418}. Along $ x $ axis, the dynamics of opposite sublattice magnetizations are symmetric, and the pumping current in this case is zero  $\langle I_{sp,x} \rangle = 0 $ . Applying STT with  $ \vec{\mu}_s^N = (1,0,0) $ only affects $\langle I_{sp,x} \rangle $, and a positive $ c_J $ drives a negative $\langle I_{sp,x} \rangle $ and hence a negative net current $\langle I_{s,x} \rangle $ (see Fig. \ref{pump-hy}(a-b)). This effect  is similar to the effect described above (Fig. \ref{cj-pumpf-h0}). However, the increase in $ \omega_{Hy} $ strongly weakens this effect, and $\langle I_{sp,x} \rangle $ and $\langle I_{s,x} \rangle $ approaches 0 for larger $ \omega_{Hy} $. To understand this phenomenon, we also analyze the change in $ {\rm Im} [\omega_{\pm}] $ induced by STT. However, this effect is negligible for lager $ \omega_{Hy} $ (not shown).

Applying STT with  polarization $ \vec{\mu}_s^N = (0,1,0) $ impacts the dynamics of the net magnetization along $ \textbf{y} $. It enhances/weakens the thermal fluctuation of the net magnetization and hence $\langle I_{sp,y} \rangle$.  As demonstrated in Fig. \ref{pump-hy}(c-d), the positive $ c_J $ decreases the $\langle I_{sp,y} \rangle$, generating a negative $\langle I_{s,y} \rangle$.  The induced $\langle I_{s,y} \rangle$ increases with the net magnetization $ m_{0,y} $ and thus magnetic field $ \omega_{Hy} $. 

After reversing the sign of $ c_J $, the current $\langle I_{s,y} \rangle$ becomes positive (Fig. \ref{pump-hy}(e)).  With further increasing of the amplitude of $ c_J $ (not shown), we observe a nonlinear variation in $\langle I_{s,y} \rangle$, where the negative $\langle I_{s,y} \rangle$ can  be  larger than the positive $\langle I_{s,y} \rangle$ under the same $ |c_J| $. This nonlinear and asymmetric variation in the net magnetization fluctuation resembles the effect in FM \cite{PhysRevB.95.020414}.  Noteworthy, as compared with the case in absence of the  magnetic field and $ \vec{\mu}_s^N \parallel \vec{n}_0 $, the  current $\langle I_{s,y} \rangle$ is much smaller  (cf. Fig. \ref{cj-pumpf-h0}), even when the magnetic field is sufficiently strong   ($ \omega_{Hy} =5.2 $ THz). Based on this observation we conclude, that when a finite net magnetization builds up along the external magnetic field, the induced spin pumping current is smaller than in the case of N$ {\rm \acute{e}} $el order vector.

\section{Conclusion}

We studied the electron-magnon spin conversion process in a HM/AFM/HM heterostructure. A charge current in the metallic layer drives a spin dynamics in the AFM via spin transfer torques (STT) effects. Two degenerate AFM magnon modes with opposite polarization are involved. Depending on the electron polarization, STT enhances one of two modes and suppresses the other. At finite temperatures, the creation/annihilation of the two magnon modes in AFM by the STT leads to a net spin pumping current. This current increases nonlinearly with the electric current density in the HM layer. 
An external magnetic field can control the conversion process, which is shown to be quite efficient and potentially  useful for designing antiferromagnetic-based spintronic devices.

\section{ACKNOWLEDGMENT}
This work was supported by the  National Natural Science Foundation of China (No. 11674400, 11374373, 11704415), DFG through SFB 762 and SFB TRR227, and the Natural Science Foundation of Hunan Province of China (No. 2018JJ3629 and 2020JJ4104).

\bibliography{Notere}

\begin{thebibliography}{44}%
\makeatletter
\providecommand \@ifxundefined [1]{%
 \@ifx{#1\undefined}
}%
\providecommand \@ifnum [1]{%
 \ifnum #1\expandafter \@firstoftwo
 \else \expandafter \@secondoftwo
 \fi
}%
\providecommand \@ifx [1]{%
 \ifx #1\expandafter \@firstoftwo
 \else \expandafter \@secondoftwo
 \fi
}%
\providecommand \natexlab [1]{#1}%
\providecommand \enquote  [1]{``#1''}%
\providecommand \bibnamefont  [1]{#1}%
\providecommand \bibfnamefont [1]{#1}%
\providecommand \citenamefont [1]{#1}%
\providecommand \href@noop [0]{\@secondoftwo}%
\providecommand \href [0]{\begingroup \@sanitize@url \@href}%
\providecommand \@href[1]{\@@startlink{#1}\@@href}%
\providecommand \@@href[1]{\endgroup#1\@@endlink}%
\providecommand \@sanitize@url [0]{\catcode `\\12\catcode `\$12\catcode
  `\&12\catcode `\#12\catcode `\^12\catcode `\_12\catcode `\%12\relax}%
\providecommand \@@startlink[1]{}%
\providecommand \@@endlink[0]{}%
\providecommand \url  [0]{\begingroup\@sanitize@url \@url }%
\providecommand \@url [1]{\endgroup\@href {#1}{\urlprefix }}%
\providecommand \urlprefix  [0]{URL }%
\providecommand \Eprint [0]{\href }%
\providecommand \doibase [0]{http://dx.doi.org/}%
\providecommand \selectlanguage [0]{\@gobble}%
\providecommand \bibinfo  [0]{\@secondoftwo}%
\providecommand \bibfield  [0]{\@secondoftwo}%
\providecommand \translation [1]{[#1]}%
\providecommand \BibitemOpen [0]{}%
\providecommand \bibitemStop [0]{}%
\providecommand \bibitemNoStop [0]{.\EOS\space}%
\providecommand \EOS [0]{\spacefactor3000\relax}%
\providecommand \BibitemShut  [1]{\csname bibitem#1\endcsname}%
\let\auto@bib@innerbib\@empty
\bibitem [{\citenamefont {\ifmmode \check{Z}\else
  \v{Z}\fi{}uti\ifmmode~\acute{c}\else \'{c}\fi{}}\ \emph
  {et~al.}(2004)\citenamefont {\ifmmode \check{Z}\else
  \v{Z}\fi{}uti\ifmmode~\acute{c}\else \'{c}\fi{}}, \citenamefont {Fabian},\
  and\ \citenamefont {Das~Sarma}}]{RevModPhys.76.323}%
  \BibitemOpen
  \bibfield  {author} {\bibinfo {author} {\bibfnamefont {I.}~\bibnamefont
  {\ifmmode \check{Z}\else \v{Z}\fi{}uti\ifmmode~\acute{c}\else \'{c}\fi{}}},
  \bibinfo {author} {\bibfnamefont {J.}~\bibnamefont {Fabian}}, \ and\ \bibinfo
  {author} {\bibfnamefont {S.}~\bibnamefont {Das~Sarma}},\ }\href {\doibase
  10.1103/RevModPhys.76.323} {\bibfield  {journal} {\bibinfo  {journal} {Rev.
  Mod. Phys.}\ }\textbf {\bibinfo {volume} {76}},\ \bibinfo {pages} {323}
  (\bibinfo {year} {2004})}\BibitemShut {NoStop}%
\bibitem [{\citenamefont {Chumak}\ \emph {et~al.}(2015)\citenamefont {Chumak},
  \citenamefont {Vasyuchka}, \citenamefont {Serga},\ and\ \citenamefont
  {Hillebrands}}]{Cchumak2015}%
  \BibitemOpen
  \bibfield  {author} {\bibinfo {author} {\bibfnamefont {A.~V.}\ \bibnamefont
  {Chumak}}, \bibinfo {author} {\bibfnamefont {V.~I.}\ \bibnamefont
  {Vasyuchka}}, \bibinfo {author} {\bibfnamefont {A.~A.}\ \bibnamefont
  {Serga}}, \ and\ \bibinfo {author} {\bibfnamefont {B.}~\bibnamefont
  {Hillebrands}},\ }\href@noop {} {\bibfield  {journal} {\bibinfo  {journal}
  {Nat. Phys.}\ }\textbf {\bibinfo {volume} {11}},\ \bibinfo {pages} {453}
  (\bibinfo {year} {2015})}\BibitemShut {NoStop}%
\bibitem [{\citenamefont {Wolf}\ \emph {et~al.}(2001)\citenamefont {Wolf},
  \citenamefont {Awschalom}, \citenamefont {Buhrman}, \citenamefont {Daughton},
  \citenamefont {von Moln{\'a}r}, \citenamefont {Roukes}, \citenamefont
  {Chtchelkanova},\ and\ \citenamefont {Treger}}]{Wolf1488}%
  \BibitemOpen
  \bibfield  {author} {\bibinfo {author} {\bibfnamefont {S.~A.}\ \bibnamefont
  {Wolf}}, \bibinfo {author} {\bibfnamefont {D.~D.}\ \bibnamefont {Awschalom}},
  \bibinfo {author} {\bibfnamefont {R.~A.}\ \bibnamefont {Buhrman}}, \bibinfo
  {author} {\bibfnamefont {J.~M.}\ \bibnamefont {Daughton}}, \bibinfo {author}
  {\bibfnamefont {S.}~\bibnamefont {von Moln{\'a}r}}, \bibinfo {author}
  {\bibfnamefont {M.~L.}\ \bibnamefont {Roukes}}, \bibinfo {author}
  {\bibfnamefont {A.~Y.}\ \bibnamefont {Chtchelkanova}}, \ and\ \bibinfo
  {author} {\bibfnamefont {D.~M.}\ \bibnamefont {Treger}},\ }\href {\doibase
  10.1126/science.1065389} {\bibfield  {journal} {\bibinfo  {journal}
  {Science}\ }\textbf {\bibinfo {volume} {294}},\ \bibinfo {pages} {1488}
  (\bibinfo {year} {2001})}\BibitemShut {NoStop}%
\bibitem [{\citenamefont {Ohno}(2010)}]{Ohno2010}%
  \BibitemOpen
  \bibfield  {author} {\bibinfo {author} {\bibfnamefont {H.}~\bibnamefont
  {Ohno}},\ }\href {\doibase 10.1038/nmat2913} {\bibfield  {journal} {\bibinfo
  {journal} {Nat. Mater.}\ }\textbf {\bibinfo {volume} {9}},\ \bibinfo {pages}
  {952} (\bibinfo {year} {2010})}\BibitemShut {NoStop}%
\bibitem [{\citenamefont {Sinova}\ \emph {et~al.}(2015)\citenamefont {Sinova},
  \citenamefont {Valenzuela}, \citenamefont {Wunderlich}, \citenamefont
  {Back},\ and\ \citenamefont {Jungwirth}}]{RevModPhys.87.1213}%
  \BibitemOpen
  \bibfield  {author} {\bibinfo {author} {\bibfnamefont {J.}~\bibnamefont
  {Sinova}}, \bibinfo {author} {\bibfnamefont {S.~O.}\ \bibnamefont
  {Valenzuela}}, \bibinfo {author} {\bibfnamefont {J.}~\bibnamefont
  {Wunderlich}}, \bibinfo {author} {\bibfnamefont {C.~H.}\ \bibnamefont
  {Back}}, \ and\ \bibinfo {author} {\bibfnamefont {T.}~\bibnamefont
  {Jungwirth}},\ }\href {\doibase 10.1103/RevModPhys.87.1213} {\bibfield
  {journal} {\bibinfo  {journal} {Rev. Mod. Phys.}\ }\textbf {\bibinfo {volume}
  {87}},\ \bibinfo {pages} {1213} (\bibinfo {year} {2015})}\BibitemShut
  {NoStop}%
\bibitem [{\citenamefont {Heinrich}\ \emph {et~al.}(2011)\citenamefont
  {Heinrich}, \citenamefont {Burrowes}, \citenamefont {Montoya}, \citenamefont
  {Kardasz}, \citenamefont {Girt}, \citenamefont {Song}, \citenamefont {Sun},\
  and\ \citenamefont {Wu}}]{PhysRevLett.107.066604}%
  \BibitemOpen
  \bibfield  {author} {\bibinfo {author} {\bibfnamefont {B.}~\bibnamefont
  {Heinrich}}, \bibinfo {author} {\bibfnamefont {C.}~\bibnamefont {Burrowes}},
  \bibinfo {author} {\bibfnamefont {E.}~\bibnamefont {Montoya}}, \bibinfo
  {author} {\bibfnamefont {B.}~\bibnamefont {Kardasz}}, \bibinfo {author}
  {\bibfnamefont {E.}~\bibnamefont {Girt}}, \bibinfo {author} {\bibfnamefont
  {Y.-Y.}\ \bibnamefont {Song}}, \bibinfo {author} {\bibfnamefont
  {Y.}~\bibnamefont {Sun}}, \ and\ \bibinfo {author} {\bibfnamefont
  {M.}~\bibnamefont {Wu}},\ }\href {\doibase 10.1103/PhysRevLett.107.066604}
  {\bibfield  {journal} {\bibinfo  {journal} {Phys. Rev. Lett.}\ }\textbf
  {\bibinfo {volume} {107}},\ \bibinfo {pages} {066604} (\bibinfo {year}
  {2011})}\BibitemShut {NoStop}%
\bibitem [{\citenamefont {Hirsch}(1999)}]{PhysRevLett.83.1834}%
  \BibitemOpen
  \bibfield  {author} {\bibinfo {author} {\bibfnamefont {J.~E.}\ \bibnamefont
  {Hirsch}},\ }\href {\doibase 10.1103/PhysRevLett.83.1834} {\bibfield
  {journal} {\bibinfo  {journal} {Phys. Rev. Lett.}\ }\textbf {\bibinfo
  {volume} {83}},\ \bibinfo {pages} {1834} (\bibinfo {year}
  {1999})}\BibitemShut {NoStop}%
\bibitem [{\citenamefont {Valenzuela}\ and\ \citenamefont
  {Tinkham}(2006)}]{Valenzuela2006}%
  \BibitemOpen
  \bibfield  {author} {\bibinfo {author} {\bibfnamefont {S.~O.}\ \bibnamefont
  {Valenzuela}}\ and\ \bibinfo {author} {\bibfnamefont {M.}~\bibnamefont
  {Tinkham}},\ }\href {\doibase 10.1038/nature04937} {\bibfield  {journal}
  {\bibinfo  {journal} {Nature}\ }\textbf {\bibinfo {volume} {442}},\ \bibinfo
  {pages} {176} (\bibinfo {year} {2006})}\BibitemShut {NoStop}%
\bibitem [{\citenamefont {Kato}\ \emph {et~al.}(2004)\citenamefont {Kato},
  \citenamefont {Myers}, \citenamefont {Gossard},\ and\ \citenamefont
  {Awschalom}}]{Kato1910}%
  \BibitemOpen
  \bibfield  {author} {\bibinfo {author} {\bibfnamefont {Y.~K.}\ \bibnamefont
  {Kato}}, \bibinfo {author} {\bibfnamefont {R.~C.}\ \bibnamefont {Myers}},
  \bibinfo {author} {\bibfnamefont {A.~C.}\ \bibnamefont {Gossard}}, \ and\
  \bibinfo {author} {\bibfnamefont {D.~D.}\ \bibnamefont {Awschalom}},\ }\href
  {\doibase 10.1126/science.1105514} {\bibfield  {journal} {\bibinfo  {journal}
  {Science}\ }\textbf {\bibinfo {volume} {306}},\ \bibinfo {pages} {1910}
  (\bibinfo {year} {2004})}\BibitemShut {NoStop}%
\bibitem [{\citenamefont {Tserkovnyak}\ \emph {et~al.}(2005)\citenamefont
  {Tserkovnyak}, \citenamefont {Brataas}, \citenamefont {Bauer},\ and\
  \citenamefont {Halperin}}]{RevModPhys.77.1375}%
  \BibitemOpen
  \bibfield  {author} {\bibinfo {author} {\bibfnamefont {Y.}~\bibnamefont
  {Tserkovnyak}}, \bibinfo {author} {\bibfnamefont {A.}~\bibnamefont
  {Brataas}}, \bibinfo {author} {\bibfnamefont {G.~E.~W.}\ \bibnamefont
  {Bauer}}, \ and\ \bibinfo {author} {\bibfnamefont {B.~I.}\ \bibnamefont
  {Halperin}},\ }\href {\doibase 10.1103/RevModPhys.77.1375} {\bibfield
  {journal} {\bibinfo  {journal} {Rev. Mod. Phys.}\ }\textbf {\bibinfo {volume}
  {77}},\ \bibinfo {pages} {1375} (\bibinfo {year} {2005})}\BibitemShut
  {NoStop}%
\bibitem [{\citenamefont {Saitoh}\ \emph {et~al.}(2006)\citenamefont {Saitoh},
  \citenamefont {Ueda}, \citenamefont {Miyajima},\ and\ \citenamefont
  {Tatara}}]{doi:10.1063/1.2199473}%
  \BibitemOpen
  \bibfield  {author} {\bibinfo {author} {\bibfnamefont {E.}~\bibnamefont
  {Saitoh}}, \bibinfo {author} {\bibfnamefont {M.}~\bibnamefont {Ueda}},
  \bibinfo {author} {\bibfnamefont {H.}~\bibnamefont {Miyajima}}, \ and\
  \bibinfo {author} {\bibfnamefont {G.}~\bibnamefont {Tatara}},\ }\href
  {\doibase 10.1063/1.2199473} {\bibfield  {journal} {\bibinfo  {journal}
  {Appl. Phys. Lett.}\ }\textbf {\bibinfo {volume} {88}},\ \bibinfo {pages}
  {182509} (\bibinfo {year} {2006})}\BibitemShut {NoStop}%
\bibitem [{\citenamefont {Sandweg}\ \emph {et~al.}(2011)\citenamefont
  {Sandweg}, \citenamefont {Kajiwara}, \citenamefont {Chumak}, \citenamefont
  {Serga}, \citenamefont {Vasyuchka}, \citenamefont {Jungfleisch},
  \citenamefont {Saitoh},\ and\ \citenamefont
  {Hillebrands}}]{PhysRevLett.106.216601}%
  \BibitemOpen
  \bibfield  {author} {\bibinfo {author} {\bibfnamefont {C.~W.}\ \bibnamefont
  {Sandweg}}, \bibinfo {author} {\bibfnamefont {Y.}~\bibnamefont {Kajiwara}},
  \bibinfo {author} {\bibfnamefont {A.~V.}\ \bibnamefont {Chumak}}, \bibinfo
  {author} {\bibfnamefont {A.~A.}\ \bibnamefont {Serga}}, \bibinfo {author}
  {\bibfnamefont {V.~I.}\ \bibnamefont {Vasyuchka}}, \bibinfo {author}
  {\bibfnamefont {M.~B.}\ \bibnamefont {Jungfleisch}}, \bibinfo {author}
  {\bibfnamefont {E.}~\bibnamefont {Saitoh}}, \ and\ \bibinfo {author}
  {\bibfnamefont {B.}~\bibnamefont {Hillebrands}},\ }\href {\doibase
  10.1103/PhysRevLett.106.216601} {\bibfield  {journal} {\bibinfo  {journal}
  {Phys. Rev. Lett.}\ }\textbf {\bibinfo {volume} {106}},\ \bibinfo {pages}
  {216601} (\bibinfo {year} {2011})}\BibitemShut {NoStop}%
\bibitem [{\citenamefont {Ando}\ \emph {et~al.}(2011)\citenamefont {Ando},
  \citenamefont {Takahashi}, \citenamefont {Ieda}, \citenamefont {Kajiwara},
  \citenamefont {Nakayama}, \citenamefont {Yoshino}, \citenamefont {Harii},
  \citenamefont {Fujikawa}, \citenamefont {Matsuo}, \citenamefont {Maekawa},\
  and\ \citenamefont {Saitoh}}]{doi:10.1063/1.3587173}%
  \BibitemOpen
  \bibfield  {author} {\bibinfo {author} {\bibfnamefont {K.}~\bibnamefont
  {Ando}}, \bibinfo {author} {\bibfnamefont {S.}~\bibnamefont {Takahashi}},
  \bibinfo {author} {\bibfnamefont {J.}~\bibnamefont {Ieda}}, \bibinfo {author}
  {\bibfnamefont {Y.}~\bibnamefont {Kajiwara}}, \bibinfo {author}
  {\bibfnamefont {H.}~\bibnamefont {Nakayama}}, \bibinfo {author}
  {\bibfnamefont {T.}~\bibnamefont {Yoshino}}, \bibinfo {author} {\bibfnamefont
  {K.}~\bibnamefont {Harii}}, \bibinfo {author} {\bibfnamefont
  {Y.}~\bibnamefont {Fujikawa}}, \bibinfo {author} {\bibfnamefont
  {M.}~\bibnamefont {Matsuo}}, \bibinfo {author} {\bibfnamefont
  {S.}~\bibnamefont {Maekawa}}, \ and\ \bibinfo {author} {\bibfnamefont
  {E.}~\bibnamefont {Saitoh}},\ }\href {\doibase 10.1063/1.3587173} {\bibfield
  {journal} {\bibinfo  {journal} {J. Appl. Phys.}\ }\textbf {\bibinfo {volume}
  {109}},\ \bibinfo {pages} {103913} (\bibinfo {year} {2011})}\BibitemShut
  {NoStop}%
\bibitem [{\citenamefont {Uchida}\ \emph {et~al.}(2010)\citenamefont {Uchida},
  \citenamefont {Xiao}, \citenamefont {Adachi}, \citenamefont {Ohe},
  \citenamefont {Takahashi}, \citenamefont {Ieda}, \citenamefont {Ota},
  \citenamefont {Kajiwara}, \citenamefont {Umezawa}, \citenamefont {Kawai},
  \citenamefont {Bauer}, \citenamefont {Maekawa},\ and\ \citenamefont
  {Saitoh}}]{Uchida2010}%
  \BibitemOpen
  \bibfield  {author} {\bibinfo {author} {\bibfnamefont {K.}~\bibnamefont
  {Uchida}}, \bibinfo {author} {\bibfnamefont {J.}~\bibnamefont {Xiao}},
  \bibinfo {author} {\bibfnamefont {H.}~\bibnamefont {Adachi}}, \bibinfo
  {author} {\bibfnamefont {J.}~\bibnamefont {Ohe}}, \bibinfo {author}
  {\bibfnamefont {S.}~\bibnamefont {Takahashi}}, \bibinfo {author}
  {\bibfnamefont {J.}~\bibnamefont {Ieda}}, \bibinfo {author} {\bibfnamefont
  {T.}~\bibnamefont {Ota}}, \bibinfo {author} {\bibfnamefont {Y.}~\bibnamefont
  {Kajiwara}}, \bibinfo {author} {\bibfnamefont {H.}~\bibnamefont {Umezawa}},
  \bibinfo {author} {\bibfnamefont {H.}~\bibnamefont {Kawai}}, \bibinfo
  {author} {\bibfnamefont {G.~E.~W.}\ \bibnamefont {Bauer}}, \bibinfo {author}
  {\bibfnamefont {S.}~\bibnamefont {Maekawa}}, \ and\ \bibinfo {author}
  {\bibfnamefont {E.}~\bibnamefont {Saitoh}},\ }\href {\doibase
  10.1038/nmat2856} {\bibfield  {journal} {\bibinfo  {journal} {Nat. Mater.}\
  }\textbf {\bibinfo {volume} {9}},\ \bibinfo {pages} {894} (\bibinfo {year}
  {2010})}\BibitemShut {NoStop}%
\bibitem [{\citenamefont {Chumak}\ \emph {et~al.}(2012)\citenamefont {Chumak},
  \citenamefont {Serga}, \citenamefont {Jungfleisch}, \citenamefont {Neb},
  \citenamefont {Bozhko}, \citenamefont {Tiberkevich},\ and\ \citenamefont
  {Hillebrands}}]{doi:10.1063/1.3689787}%
  \BibitemOpen
  \bibfield  {author} {\bibinfo {author} {\bibfnamefont {A.~V.}\ \bibnamefont
  {Chumak}}, \bibinfo {author} {\bibfnamefont {A.~A.}\ \bibnamefont {Serga}},
  \bibinfo {author} {\bibfnamefont {M.~B.}\ \bibnamefont {Jungfleisch}},
  \bibinfo {author} {\bibfnamefont {R.}~\bibnamefont {Neb}}, \bibinfo {author}
  {\bibfnamefont {D.~A.}\ \bibnamefont {Bozhko}}, \bibinfo {author}
  {\bibfnamefont {V.~S.}\ \bibnamefont {Tiberkevich}}, \ and\ \bibinfo {author}
  {\bibfnamefont {B.}~\bibnamefont {Hillebrands}},\ }\href {\doibase
  10.1063/1.3689787} {\bibfield  {journal} {\bibinfo  {journal} {Appl. Phys.
  Lett.}\ }\textbf {\bibinfo {volume} {100}},\ \bibinfo {pages} {082405}
  (\bibinfo {year} {2012})}\BibitemShut {NoStop}%
\bibitem [{\citenamefont {Kajiwara}\ \emph {et~al.}(2010)\citenamefont
  {Kajiwara}, \citenamefont {Harii}, \citenamefont {Takahashi}, \citenamefont
  {Ohe}, \citenamefont {Uchida}, \citenamefont {Mizuguchi}, \citenamefont
  {Umezawa}, \citenamefont {Kawai}, \citenamefont {Ando}, \citenamefont
  {Takanashi}, \citenamefont {Maekawa},\ and\ \citenamefont
  {Saitoh}}]{Kajiwara2010}%
  \BibitemOpen
  \bibfield  {author} {\bibinfo {author} {\bibfnamefont {Y.}~\bibnamefont
  {Kajiwara}}, \bibinfo {author} {\bibfnamefont {K.}~\bibnamefont {Harii}},
  \bibinfo {author} {\bibfnamefont {S.}~\bibnamefont {Takahashi}}, \bibinfo
  {author} {\bibfnamefont {J.}~\bibnamefont {Ohe}}, \bibinfo {author}
  {\bibfnamefont {K.}~\bibnamefont {Uchida}}, \bibinfo {author} {\bibfnamefont
  {M.}~\bibnamefont {Mizuguchi}}, \bibinfo {author} {\bibfnamefont
  {H.}~\bibnamefont {Umezawa}}, \bibinfo {author} {\bibfnamefont
  {H.}~\bibnamefont {Kawai}}, \bibinfo {author} {\bibfnamefont
  {K.}~\bibnamefont {Ando}}, \bibinfo {author} {\bibfnamefont {K.}~\bibnamefont
  {Takanashi}}, \bibinfo {author} {\bibfnamefont {S.}~\bibnamefont {Maekawa}},
  \ and\ \bibinfo {author} {\bibfnamefont {E.}~\bibnamefont {Saitoh}},\ }\href
  {\doibase 10.1038/nature08876} {\bibfield  {journal} {\bibinfo  {journal}
  {Nature}\ }\textbf {\bibinfo {volume} {464}},\ \bibinfo {pages} {262}
  (\bibinfo {year} {2010})}\BibitemShut {NoStop}%
\bibitem [{\citenamefont {Wang}\ \emph {et~al.}(2017)\citenamefont {Wang},
  \citenamefont {Li}, \citenamefont {Zhou}, \citenamefont {Nie}, \citenamefont
  {Xia}, \citenamefont {Zeng}, \citenamefont {Chotorlishvili}, \citenamefont
  {Berakdar},\ and\ \citenamefont {Guo}}]{PhysRevB.95.020414}%
  \BibitemOpen
  \bibfield  {author} {\bibinfo {author} {\bibfnamefont {X.-g.}\ \bibnamefont
  {Wang}}, \bibinfo {author} {\bibfnamefont {Z.-x.}\ \bibnamefont {Li}},
  \bibinfo {author} {\bibfnamefont {Z.-w.}\ \bibnamefont {Zhou}}, \bibinfo
  {author} {\bibfnamefont {Y.-z.}\ \bibnamefont {Nie}}, \bibinfo {author}
  {\bibfnamefont {Q.-l.}\ \bibnamefont {Xia}}, \bibinfo {author} {\bibfnamefont
  {Z.-m.}\ \bibnamefont {Zeng}}, \bibinfo {author} {\bibfnamefont
  {L.}~\bibnamefont {Chotorlishvili}}, \bibinfo {author} {\bibfnamefont
  {J.}~\bibnamefont {Berakdar}}, \ and\ \bibinfo {author} {\bibfnamefont
  {G.-h.}\ \bibnamefont {Guo}},\ }\href {\doibase 10.1103/PhysRevB.95.020414}
  {\bibfield  {journal} {\bibinfo  {journal} {Phys. Rev. B}\ }\textbf {\bibinfo
  {volume} {95}},\ \bibinfo {pages} {020414} (\bibinfo {year}
  {2017})}\BibitemShut {NoStop}%
\bibitem [{\citenamefont {Chotorlishvili}\ \emph {et~al.}(2019)\citenamefont
  {Chotorlishvili}, \citenamefont {Toklikishvili}, \citenamefont {Wang},
  \citenamefont {Dugaev}, \citenamefont {Barna\ifmmode~\acute{s}\else
  \'{s}\fi{}},\ and\ \citenamefont {Berakdar}}]{PhysRevB.99.024410}%
  \BibitemOpen
  \bibfield  {author} {\bibinfo {author} {\bibfnamefont {L.}~\bibnamefont
  {Chotorlishvili}}, \bibinfo {author} {\bibfnamefont {Z.}~\bibnamefont
  {Toklikishvili}}, \bibinfo {author} {\bibfnamefont {X.-G.}\ \bibnamefont
  {Wang}}, \bibinfo {author} {\bibfnamefont {V.~K.}\ \bibnamefont {Dugaev}},
  \bibinfo {author} {\bibfnamefont {J.}~\bibnamefont
  {Barna\ifmmode~\acute{s}\else \'{s}\fi{}}}, \ and\ \bibinfo {author}
  {\bibfnamefont {J.}~\bibnamefont {Berakdar}},\ }\href {\doibase
  10.1103/PhysRevB.99.024410} {\bibfield  {journal} {\bibinfo  {journal} {Phys.
  Rev. B}\ }\textbf {\bibinfo {volume} {99}},\ \bibinfo {pages} {024410}
  (\bibinfo {year} {2019})}\BibitemShut {NoStop}%
\bibitem [{\citenamefont {Wang}\ \emph {et~al.}(2012)\citenamefont {Wang},
  \citenamefont {Guo}, \citenamefont {Nie}, \citenamefont {Zhang},\ and\
  \citenamefont {Li}}]{PhysRevB.86.054445}%
  \BibitemOpen
  \bibfield  {author} {\bibinfo {author} {\bibfnamefont {X.-g.}\ \bibnamefont
  {Wang}}, \bibinfo {author} {\bibfnamefont {G.-h.}\ \bibnamefont {Guo}},
  \bibinfo {author} {\bibfnamefont {Y.-z.}\ \bibnamefont {Nie}}, \bibinfo
  {author} {\bibfnamefont {G.-f.}\ \bibnamefont {Zhang}}, \ and\ \bibinfo
  {author} {\bibfnamefont {Z.-x.}\ \bibnamefont {Li}},\ }\href {\doibase
  10.1103/PhysRevB.86.054445} {\bibfield  {journal} {\bibinfo  {journal} {Phys.
  Rev. B}\ }\textbf {\bibinfo {volume} {86}},\ \bibinfo {pages} {054445}
  (\bibinfo {year} {2012})}\BibitemShut {NoStop}%
\bibitem [{\citenamefont {Jungwirth}\ \emph {et~al.}(2016)\citenamefont
  {Jungwirth}, \citenamefont {Marti}, \citenamefont {Wadley},\ and\
  \citenamefont {Wunderlich}}]{Jungwirth2016}%
  \BibitemOpen
  \bibfield  {author} {\bibinfo {author} {\bibfnamefont {T.}~\bibnamefont
  {Jungwirth}}, \bibinfo {author} {\bibfnamefont {X.}~\bibnamefont {Marti}},
  \bibinfo {author} {\bibfnamefont {P.}~\bibnamefont {Wadley}}, \ and\ \bibinfo
  {author} {\bibfnamefont {J.}~\bibnamefont {Wunderlich}},\ }\href {\doibase
  10.1038/nnano.2016.18} {\bibfield  {journal} {\bibinfo  {journal} {Nat.
  Nanotechnol.}\ }\textbf {\bibinfo {volume} {11}},\ \bibinfo {pages} {231}
  (\bibinfo {year} {2016})}\BibitemShut {NoStop}%
\bibitem [{\citenamefont {Gomonay}\ and\ \citenamefont
  {Loktev}(2014)}]{doi:10.1063/1.4862467}%
  \BibitemOpen
  \bibfield  {author} {\bibinfo {author} {\bibfnamefont {E.~V.}\ \bibnamefont
  {Gomonay}}\ and\ \bibinfo {author} {\bibfnamefont {V.~M.}\ \bibnamefont
  {Loktev}},\ }\href {\doibase 10.1063/1.4862467} {\bibfield  {journal}
  {\bibinfo  {journal} {Low Temp. Phys.}\ }\textbf {\bibinfo {volume} {40}},\
  \bibinfo {pages} {17} (\bibinfo {year} {2014})}\BibitemShut {NoStop}%
\bibitem [{\citenamefont {Baltz}\ \emph {et~al.}(2018)\citenamefont {Baltz},
  \citenamefont {Manchon}, \citenamefont {Tsoi}, \citenamefont {Moriyama},
  \citenamefont {Ono},\ and\ \citenamefont
  {Tserkovnyak}}]{RevModPhys.90.015005}%
  \BibitemOpen
  \bibfield  {author} {\bibinfo {author} {\bibfnamefont {V.}~\bibnamefont
  {Baltz}}, \bibinfo {author} {\bibfnamefont {A.}~\bibnamefont {Manchon}},
  \bibinfo {author} {\bibfnamefont {M.}~\bibnamefont {Tsoi}}, \bibinfo {author}
  {\bibfnamefont {T.}~\bibnamefont {Moriyama}}, \bibinfo {author}
  {\bibfnamefont {T.}~\bibnamefont {Ono}}, \ and\ \bibinfo {author}
  {\bibfnamefont {Y.}~\bibnamefont {Tserkovnyak}},\ }\href {\doibase
  10.1103/RevModPhys.90.015005} {\bibfield  {journal} {\bibinfo  {journal}
  {Rev. Mod. Phys.}\ }\textbf {\bibinfo {volume} {90}},\ \bibinfo {pages}
  {015005} (\bibinfo {year} {2018})}\BibitemShut {NoStop}%
\bibitem [{\citenamefont {Gomonay}\ \emph {et~al.}(2018)\citenamefont
  {Gomonay}, \citenamefont {Baltz}, \citenamefont {Brataas},\ and\
  \citenamefont {Tserkovnyak}}]{Gomonay2018}%
  \BibitemOpen
  \bibfield  {author} {\bibinfo {author} {\bibfnamefont {O.}~\bibnamefont
  {Gomonay}}, \bibinfo {author} {\bibfnamefont {V.}~\bibnamefont {Baltz}},
  \bibinfo {author} {\bibfnamefont {A.}~\bibnamefont {Brataas}}, \ and\
  \bibinfo {author} {\bibfnamefont {Y.}~\bibnamefont {Tserkovnyak}},\ }\href
  {\doibase 10.1038/s41567-018-0049-4} {\bibfield  {journal} {\bibinfo
  {journal} {Nat. Phys.}\ }\textbf {\bibinfo {volume} {14}},\ \bibinfo {pages}
  {213} (\bibinfo {year} {2018})}\BibitemShut {NoStop}%
\bibitem [{\citenamefont {Jungfleisch}\ \emph {et~al.}(2018)\citenamefont
  {Jungfleisch}, \citenamefont {Zhang},\ and\ \citenamefont
  {Hoffmann}}]{JUNGFLEISCH2018865}%
  \BibitemOpen
  \bibfield  {author} {\bibinfo {author} {\bibfnamefont {M.~B.}\ \bibnamefont
  {Jungfleisch}}, \bibinfo {author} {\bibfnamefont {W.}~\bibnamefont {Zhang}},
  \ and\ \bibinfo {author} {\bibfnamefont {A.}~\bibnamefont {Hoffmann}},\
  }\href {\doibase https://doi.org/10.1016/j.physleta.2018.01.008} {\bibfield
  {journal} {\bibinfo  {journal} {Phys. Lett. A}\ }\textbf {\bibinfo {volume}
  {382}},\ \bibinfo {pages} {865 } (\bibinfo {year} {2018})}\BibitemShut
  {NoStop}%
\bibitem [{\citenamefont {Park}\ \emph {et~al.}(2011)\citenamefont {Park},
  \citenamefont {Wunderlich}, \citenamefont {Mart{\'i}}, \citenamefont
  {Hol{\'y}}, \citenamefont {Kurosaki}, \citenamefont {Yamada}, \citenamefont
  {Yamamoto}, \citenamefont {Nishide}, \citenamefont {Hayakawa}, \citenamefont
  {Takahashi}, \citenamefont {Shick},\ and\ \citenamefont
  {Jungwirth}}]{Park2011}%
  \BibitemOpen
  \bibfield  {author} {\bibinfo {author} {\bibfnamefont {B.~G.}\ \bibnamefont
  {Park}}, \bibinfo {author} {\bibfnamefont {J.}~\bibnamefont {Wunderlich}},
  \bibinfo {author} {\bibfnamefont {X.}~\bibnamefont {Mart{\'i}}}, \bibinfo
  {author} {\bibfnamefont {V.}~\bibnamefont {Hol{\'y}}}, \bibinfo {author}
  {\bibfnamefont {Y.}~\bibnamefont {Kurosaki}}, \bibinfo {author}
  {\bibfnamefont {M.}~\bibnamefont {Yamada}}, \bibinfo {author} {\bibfnamefont
  {H.}~\bibnamefont {Yamamoto}}, \bibinfo {author} {\bibfnamefont
  {A.}~\bibnamefont {Nishide}}, \bibinfo {author} {\bibfnamefont
  {J.}~\bibnamefont {Hayakawa}}, \bibinfo {author} {\bibfnamefont
  {H.}~\bibnamefont {Takahashi}}, \bibinfo {author} {\bibfnamefont {A.~B.}\
  \bibnamefont {Shick}}, \ and\ \bibinfo {author} {\bibfnamefont
  {T.}~\bibnamefont {Jungwirth}},\ }\href {\doibase 10.1038/nmat2983}
  {\bibfield  {journal} {\bibinfo  {journal} {Nature Materials}\ }\textbf
  {\bibinfo {volume} {10}},\ \bibinfo {pages} {347} (\bibinfo {year}
  {2011})}\BibitemShut {NoStop}%
\bibitem [{\citenamefont {Rezende}\ \emph
  {et~al.}(2016{\natexlab{a}})\citenamefont {Rezende}, \citenamefont
  {Rodr\'{\i}guez-Su\'arez},\ and\ \citenamefont
  {Azevedo}}]{PhysRevB.93.014425}%
  \BibitemOpen
  \bibfield  {author} {\bibinfo {author} {\bibfnamefont {S.~M.}\ \bibnamefont
  {Rezende}}, \bibinfo {author} {\bibfnamefont {R.~L.}\ \bibnamefont
  {Rodr\'{\i}guez-Su\'arez}}, \ and\ \bibinfo {author} {\bibfnamefont
  {A.}~\bibnamefont {Azevedo}},\ }\href {\doibase 10.1103/PhysRevB.93.014425}
  {\bibfield  {journal} {\bibinfo  {journal} {Phys. Rev. B}\ }\textbf {\bibinfo
  {volume} {93}},\ \bibinfo {pages} {014425} (\bibinfo {year}
  {2016}{\natexlab{a}})}\BibitemShut {NoStop}%
\bibitem [{\citenamefont {Rezende}\ \emph
  {et~al.}(2016{\natexlab{b}})\citenamefont {Rezende}, \citenamefont
  {Rodr\'{\i}guez-Su\'arez},\ and\ \citenamefont
  {Azevedo}}]{PhysRevB.93.054412}%
  \BibitemOpen
  \bibfield  {author} {\bibinfo {author} {\bibfnamefont {S.~M.}\ \bibnamefont
  {Rezende}}, \bibinfo {author} {\bibfnamefont {R.~L.}\ \bibnamefont
  {Rodr\'{\i}guez-Su\'arez}}, \ and\ \bibinfo {author} {\bibfnamefont
  {A.}~\bibnamefont {Azevedo}},\ }\href {\doibase 10.1103/PhysRevB.93.054412}
  {\bibfield  {journal} {\bibinfo  {journal} {Phys. Rev. B}\ }\textbf {\bibinfo
  {volume} {93}},\ \bibinfo {pages} {054412} (\bibinfo {year}
  {2016}{\natexlab{b}})}\BibitemShut {NoStop}%
\bibitem [{\citenamefont {Wadley}\ \emph {et~al.}(2016)\citenamefont {Wadley},
  \citenamefont {Howells}, \citenamefont {{\v Z}elezn{\'y}}, \citenamefont
  {Andrews}, \citenamefont {Hills}, \citenamefont {Campion}, \citenamefont
  {Nov{\'a}k}, \citenamefont {Olejn{\'\i}k}, \citenamefont {Maccherozzi},
  \citenamefont {Dhesi}, \citenamefont {Martin}, \citenamefont {Wagner},
  \citenamefont {Wunderlich}, \citenamefont {Freimuth}, \citenamefont
  {Mokrousov}, \citenamefont {Kune{\v s}}, \citenamefont {Chauhan},
  \citenamefont {Grzybowski}, \citenamefont {Rushforth}, \citenamefont
  {Edmonds}, \citenamefont {Gallagher},\ and\ \citenamefont
  {Jungwirth}}]{Wadley587}%
  \BibitemOpen
  \bibfield  {author} {\bibinfo {author} {\bibfnamefont {P.}~\bibnamefont
  {Wadley}}, \bibinfo {author} {\bibfnamefont {B.}~\bibnamefont {Howells}},
  \bibinfo {author} {\bibfnamefont {J.}~\bibnamefont {{\v Z}elezn{\'y}}},
  \bibinfo {author} {\bibfnamefont {C.}~\bibnamefont {Andrews}}, \bibinfo
  {author} {\bibfnamefont {V.}~\bibnamefont {Hills}}, \bibinfo {author}
  {\bibfnamefont {R.~P.}\ \bibnamefont {Campion}}, \bibinfo {author}
  {\bibfnamefont {V.}~\bibnamefont {Nov{\'a}k}}, \bibinfo {author}
  {\bibfnamefont {K.}~\bibnamefont {Olejn{\'\i}k}}, \bibinfo {author}
  {\bibfnamefont {F.}~\bibnamefont {Maccherozzi}}, \bibinfo {author}
  {\bibfnamefont {S.~S.}\ \bibnamefont {Dhesi}}, \bibinfo {author}
  {\bibfnamefont {S.~Y.}\ \bibnamefont {Martin}}, \bibinfo {author}
  {\bibfnamefont {T.}~\bibnamefont {Wagner}}, \bibinfo {author} {\bibfnamefont
  {J.}~\bibnamefont {Wunderlich}}, \bibinfo {author} {\bibfnamefont
  {F.}~\bibnamefont {Freimuth}}, \bibinfo {author} {\bibfnamefont
  {Y.}~\bibnamefont {Mokrousov}}, \bibinfo {author} {\bibfnamefont
  {J.}~\bibnamefont {Kune{\v s}}}, \bibinfo {author} {\bibfnamefont {J.~S.}\
  \bibnamefont {Chauhan}}, \bibinfo {author} {\bibfnamefont {M.~J.}\
  \bibnamefont {Grzybowski}}, \bibinfo {author} {\bibfnamefont {A.~W.}\
  \bibnamefont {Rushforth}}, \bibinfo {author} {\bibfnamefont {K.~W.}\
  \bibnamefont {Edmonds}}, \bibinfo {author} {\bibfnamefont {B.~L.}\
  \bibnamefont {Gallagher}}, \ and\ \bibinfo {author} {\bibfnamefont
  {T.}~\bibnamefont {Jungwirth}},\ }\href {\doibase 10.1126/science.aab1031}
  {\bibfield  {journal} {\bibinfo  {journal} {Science}\ }\textbf {\bibinfo
  {volume} {351}},\ \bibinfo {pages} {587} (\bibinfo {year}
  {2016})}\BibitemShut {NoStop}%
\bibitem [{\citenamefont {Wu}\ \emph {et~al.}(2016)\citenamefont {Wu},
  \citenamefont {Zhang}, \citenamefont {KC}, \citenamefont {Borisov},
  \citenamefont {Pearson}, \citenamefont {Jiang}, \citenamefont {Lederman},
  \citenamefont {Hoffmann},\ and\ \citenamefont
  {Bhattacharya}}]{PhysRevLett.116.097204}%
  \BibitemOpen
  \bibfield  {author} {\bibinfo {author} {\bibfnamefont {S.~M.}\ \bibnamefont
  {Wu}}, \bibinfo {author} {\bibfnamefont {W.}~\bibnamefont {Zhang}}, \bibinfo
  {author} {\bibfnamefont {A.}~\bibnamefont {KC}}, \bibinfo {author}
  {\bibfnamefont {P.}~\bibnamefont {Borisov}}, \bibinfo {author} {\bibfnamefont
  {J.~E.}\ \bibnamefont {Pearson}}, \bibinfo {author} {\bibfnamefont {J.~S.}\
  \bibnamefont {Jiang}}, \bibinfo {author} {\bibfnamefont {D.}~\bibnamefont
  {Lederman}}, \bibinfo {author} {\bibfnamefont {A.}~\bibnamefont {Hoffmann}},
  \ and\ \bibinfo {author} {\bibfnamefont {A.}~\bibnamefont {Bhattacharya}},\
  }\href {\doibase 10.1103/PhysRevLett.116.097204} {\bibfield  {journal}
  {\bibinfo  {journal} {Phys. Rev. Lett.}\ }\textbf {\bibinfo {volume} {116}},\
  \bibinfo {pages} {097204} (\bibinfo {year} {2016})}\BibitemShut {NoStop}%
\bibitem [{\citenamefont {Chen}\ \emph {et~al.}(2018)\citenamefont {Chen},
  \citenamefont {Zarzuela}, \citenamefont {Zhang}, \citenamefont {Song},
  \citenamefont {Zhou}, \citenamefont {Shi}, \citenamefont {Li}, \citenamefont
  {Zhou}, \citenamefont {Jiang}, \citenamefont {Pan},\ and\ \citenamefont
  {Tserkovnyak}}]{PhysRevLett.120.207204}%
  \BibitemOpen
  \bibfield  {author} {\bibinfo {author} {\bibfnamefont {X.~Z.}\ \bibnamefont
  {Chen}}, \bibinfo {author} {\bibfnamefont {R.}~\bibnamefont {Zarzuela}},
  \bibinfo {author} {\bibfnamefont {J.}~\bibnamefont {Zhang}}, \bibinfo
  {author} {\bibfnamefont {C.}~\bibnamefont {Song}}, \bibinfo {author}
  {\bibfnamefont {X.~F.}\ \bibnamefont {Zhou}}, \bibinfo {author}
  {\bibfnamefont {G.~Y.}\ \bibnamefont {Shi}}, \bibinfo {author} {\bibfnamefont
  {F.}~\bibnamefont {Li}}, \bibinfo {author} {\bibfnamefont {H.~A.}\
  \bibnamefont {Zhou}}, \bibinfo {author} {\bibfnamefont {W.~J.}\ \bibnamefont
  {Jiang}}, \bibinfo {author} {\bibfnamefont {F.}~\bibnamefont {Pan}}, \ and\
  \bibinfo {author} {\bibfnamefont {Y.}~\bibnamefont {Tserkovnyak}},\ }\href
  {\doibase 10.1103/PhysRevLett.120.207204} {\bibfield  {journal} {\bibinfo
  {journal} {Phys. Rev. Lett.}\ }\textbf {\bibinfo {volume} {120}},\ \bibinfo
  {pages} {207204} (\bibinfo {year} {2018})}\BibitemShut {NoStop}%
\bibitem [{\citenamefont {Wang}\ \emph {et~al.}(2014)\citenamefont {Wang},
  \citenamefont {Du}, \citenamefont {Hammel},\ and\ \citenamefont
  {Yang}}]{PhysRevLett.113.097202}%
  \BibitemOpen
  \bibfield  {author} {\bibinfo {author} {\bibfnamefont {H.}~\bibnamefont
  {Wang}}, \bibinfo {author} {\bibfnamefont {C.}~\bibnamefont {Du}}, \bibinfo
  {author} {\bibfnamefont {P.~C.}\ \bibnamefont {Hammel}}, \ and\ \bibinfo
  {author} {\bibfnamefont {F.}~\bibnamefont {Yang}},\ }\href {\doibase
  10.1103/PhysRevLett.113.097202} {\bibfield  {journal} {\bibinfo  {journal}
  {Phys. Rev. Lett.}\ }\textbf {\bibinfo {volume} {113}},\ \bibinfo {pages}
  {097202} (\bibinfo {year} {2014})}\BibitemShut {NoStop}%
\bibitem [{\citenamefont {Lin}\ \emph {et~al.}(2016)\citenamefont {Lin},
  \citenamefont {Chen}, \citenamefont {Zhang},\ and\ \citenamefont
  {Chien}}]{PhysRevLett.116.186601}%
  \BibitemOpen
  \bibfield  {author} {\bibinfo {author} {\bibfnamefont {W.}~\bibnamefont
  {Lin}}, \bibinfo {author} {\bibfnamefont {K.}~\bibnamefont {Chen}}, \bibinfo
  {author} {\bibfnamefont {S.}~\bibnamefont {Zhang}}, \ and\ \bibinfo {author}
  {\bibfnamefont {C.~L.}\ \bibnamefont {Chien}},\ }\href {\doibase
  10.1103/PhysRevLett.116.186601} {\bibfield  {journal} {\bibinfo  {journal}
  {Phys. Rev. Lett.}\ }\textbf {\bibinfo {volume} {116}},\ \bibinfo {pages}
  {186601} (\bibinfo {year} {2016})}\BibitemShut {NoStop}%
\bibitem [{\citenamefont {Moriyama}\ \emph {et~al.}(2015)\citenamefont
  {Moriyama}, \citenamefont {Takei}, \citenamefont {Nagata}, \citenamefont
  {Yoshimura}, \citenamefont {Matsuzaki}, \citenamefont {Terashima},
  \citenamefont {Tserkovnyak},\ and\ \citenamefont
  {Ono}}]{doi:10.1063/1.4918990}%
  \BibitemOpen
  \bibfield  {author} {\bibinfo {author} {\bibfnamefont {T.}~\bibnamefont
  {Moriyama}}, \bibinfo {author} {\bibfnamefont {S.}~\bibnamefont {Takei}},
  \bibinfo {author} {\bibfnamefont {M.}~\bibnamefont {Nagata}}, \bibinfo
  {author} {\bibfnamefont {Y.}~\bibnamefont {Yoshimura}}, \bibinfo {author}
  {\bibfnamefont {N.}~\bibnamefont {Matsuzaki}}, \bibinfo {author}
  {\bibfnamefont {T.}~\bibnamefont {Terashima}}, \bibinfo {author}
  {\bibfnamefont {Y.}~\bibnamefont {Tserkovnyak}}, \ and\ \bibinfo {author}
  {\bibfnamefont {T.}~\bibnamefont {Ono}},\ }\href {\doibase 10.1063/1.4918990}
  {\bibfield  {journal} {\bibinfo  {journal} {Appl. Phys. Lett.}\ }\textbf
  {\bibinfo {volume} {106}},\ \bibinfo {pages} {162406} (\bibinfo {year}
  {2015})}\BibitemShut {NoStop}%
\bibitem [{\citenamefont {Li}\ \emph {et~al.}(2020)\citenamefont {Li},
  \citenamefont {Wilson}, \citenamefont {Cheng}, \citenamefont {Lohmann},
  \citenamefont {Kavand}, \citenamefont {Yuan}, \citenamefont {Aldosary},
  \citenamefont {Agladze}, \citenamefont {Wei}, \citenamefont {Sherwin},\ and\
  \citenamefont {Shi}}]{Li2020}%
  \BibitemOpen
  \bibfield  {author} {\bibinfo {author} {\bibfnamefont {J.}~\bibnamefont
  {Li}}, \bibinfo {author} {\bibfnamefont {C.~B.}\ \bibnamefont {Wilson}},
  \bibinfo {author} {\bibfnamefont {R.}~\bibnamefont {Cheng}}, \bibinfo
  {author} {\bibfnamefont {M.}~\bibnamefont {Lohmann}}, \bibinfo {author}
  {\bibfnamefont {M.}~\bibnamefont {Kavand}}, \bibinfo {author} {\bibfnamefont
  {W.}~\bibnamefont {Yuan}}, \bibinfo {author} {\bibfnamefont {M.}~\bibnamefont
  {Aldosary}}, \bibinfo {author} {\bibfnamefont {N.}~\bibnamefont {Agladze}},
  \bibinfo {author} {\bibfnamefont {P.}~\bibnamefont {Wei}}, \bibinfo {author}
  {\bibfnamefont {M.~S.}\ \bibnamefont {Sherwin}}, \ and\ \bibinfo {author}
  {\bibfnamefont {J.}~\bibnamefont {Shi}},\ }\href {\doibase
  10.1038/s41586-020-1950-4} {\bibfield  {journal} {\bibinfo  {journal}
  {Nature}\ }\textbf {\bibinfo {volume} {578}},\ \bibinfo {pages} {70}
  (\bibinfo {year} {2020})}\BibitemShut {NoStop}%
\bibitem [{\citenamefont {Vaidya}\ \emph {et~al.}(2020)\citenamefont {Vaidya},
  \citenamefont {Morley}, \citenamefont {van Tol}, \citenamefont {Liu},
  \citenamefont {Cheng}, \citenamefont {Brataas}, \citenamefont {Lederman},\
  and\ \citenamefont {del Barco}}]{Vaidya160}%
  \BibitemOpen
  \bibfield  {author} {\bibinfo {author} {\bibfnamefont {P.}~\bibnamefont
  {Vaidya}}, \bibinfo {author} {\bibfnamefont {S.~A.}\ \bibnamefont {Morley}},
  \bibinfo {author} {\bibfnamefont {J.}~\bibnamefont {van Tol}}, \bibinfo
  {author} {\bibfnamefont {Y.}~\bibnamefont {Liu}}, \bibinfo {author}
  {\bibfnamefont {R.}~\bibnamefont {Cheng}}, \bibinfo {author} {\bibfnamefont
  {A.}~\bibnamefont {Brataas}}, \bibinfo {author} {\bibfnamefont
  {D.}~\bibnamefont {Lederman}}, \ and\ \bibinfo {author} {\bibfnamefont
  {E.}~\bibnamefont {del Barco}},\ }\href {\doibase 10.1126/science.aaz4247}
  {\bibfield  {journal} {\bibinfo  {journal} {Science}\ }\textbf {\bibinfo
  {volume} {368}},\ \bibinfo {pages} {160} (\bibinfo {year}
  {2020})}\BibitemShut {NoStop}%
\bibitem [{\citenamefont {Lebrun}\ \emph {et~al.}(2018)\citenamefont {Lebrun},
  \citenamefont {Ross}, \citenamefont {Bender}, \citenamefont {Qaiumzadeh},
  \citenamefont {Baldrati}, \citenamefont {Cramer}, \citenamefont {Brataas},
  \citenamefont {Duine},\ and\ \citenamefont {Kl{\"a}ui}}]{Lebrun2018}%
  \BibitemOpen
  \bibfield  {author} {\bibinfo {author} {\bibfnamefont {R.}~\bibnamefont
  {Lebrun}}, \bibinfo {author} {\bibfnamefont {A.}~\bibnamefont {Ross}},
  \bibinfo {author} {\bibfnamefont {S.~A.}\ \bibnamefont {Bender}}, \bibinfo
  {author} {\bibfnamefont {A.}~\bibnamefont {Qaiumzadeh}}, \bibinfo {author}
  {\bibfnamefont {L.}~\bibnamefont {Baldrati}}, \bibinfo {author}
  {\bibfnamefont {J.}~\bibnamefont {Cramer}}, \bibinfo {author} {\bibfnamefont
  {A.}~\bibnamefont {Brataas}}, \bibinfo {author} {\bibfnamefont {R.~A.}\
  \bibnamefont {Duine}}, \ and\ \bibinfo {author} {\bibfnamefont
  {M.}~\bibnamefont {Kl{\"a}ui}},\ }\href {\doibase 10.1038/s41586-018-0490-7}
  {\bibfield  {journal} {\bibinfo  {journal} {Nature}\ }\textbf {\bibinfo
  {volume} {561}},\ \bibinfo {pages} {222} (\bibinfo {year}
  {2018})}\BibitemShut {NoStop}%
\bibitem [{\citenamefont {Dabrowski}\ \emph {et~al.}(2020)\citenamefont
  {Dabrowski}, \citenamefont {Nakano}, \citenamefont {Burn}, \citenamefont
  {Frisk}, \citenamefont {Newman}, \citenamefont {Klewe}, \citenamefont {Li},
  \citenamefont {Yang}, \citenamefont {Shafer}, \citenamefont {Arenholz},
  \citenamefont {Hesjedal}, \citenamefont {van~der Laan}, \citenamefont {Qiu},\
  and\ \citenamefont {Hicken}}]{PhysRevLett.124.217201}%
  \BibitemOpen
  \bibfield  {author} {\bibinfo {author} {\bibfnamefont {M.}~\bibnamefont
  {Dabrowski}}, \bibinfo {author} {\bibfnamefont {T.}~\bibnamefont {Nakano}},
  \bibinfo {author} {\bibfnamefont {D.~M.}\ \bibnamefont {Burn}}, \bibinfo
  {author} {\bibfnamefont {A.}~\bibnamefont {Frisk}}, \bibinfo {author}
  {\bibfnamefont {D.~G.}\ \bibnamefont {Newman}}, \bibinfo {author}
  {\bibfnamefont {C.}~\bibnamefont {Klewe}}, \bibinfo {author} {\bibfnamefont
  {Q.}~\bibnamefont {Li}}, \bibinfo {author} {\bibfnamefont {M.}~\bibnamefont
  {Yang}}, \bibinfo {author} {\bibfnamefont {P.}~\bibnamefont {Shafer}},
  \bibinfo {author} {\bibfnamefont {E.}~\bibnamefont {Arenholz}}, \bibinfo
  {author} {\bibfnamefont {T.}~\bibnamefont {Hesjedal}}, \bibinfo {author}
  {\bibfnamefont {G.}~\bibnamefont {van~der Laan}}, \bibinfo {author}
  {\bibfnamefont {Z.~Q.}\ \bibnamefont {Qiu}}, \ and\ \bibinfo {author}
  {\bibfnamefont {R.~J.}\ \bibnamefont {Hicken}},\ }\href {\doibase
  10.1103/PhysRevLett.124.217201} {\bibfield  {journal} {\bibinfo  {journal}
  {Phys. Rev. Lett.}\ }\textbf {\bibinfo {volume} {124}},\ \bibinfo {pages}
  {217201} (\bibinfo {year} {2020})}\BibitemShut {NoStop}%
\bibitem [{\citenamefont {Cheng}\ \emph {et~al.}(2016)\citenamefont {Cheng},
  \citenamefont {Xiao},\ and\ \citenamefont
  {Brataas}}]{PhysRevLett.116.207603}%
  \BibitemOpen
  \bibfield  {author} {\bibinfo {author} {\bibfnamefont {R.}~\bibnamefont
  {Cheng}}, \bibinfo {author} {\bibfnamefont {D.}~\bibnamefont {Xiao}}, \ and\
  \bibinfo {author} {\bibfnamefont {A.}~\bibnamefont {Brataas}},\ }\href
  {\doibase 10.1103/PhysRevLett.116.207603} {\bibfield  {journal} {\bibinfo
  {journal} {Phys. Rev. Lett.}\ }\textbf {\bibinfo {volume} {116}},\ \bibinfo
  {pages} {207603} (\bibinfo {year} {2016})}\BibitemShut {NoStop}%
\bibitem [{\citenamefont {Johansen}\ and\ \citenamefont
  {Brataas}(2017)}]{PhysRevB.95.220408}%
  \BibitemOpen
  \bibfield  {author} {\bibinfo {author} {\bibfnamefont {O.}~\bibnamefont
  {Johansen}}\ and\ \bibinfo {author} {\bibfnamefont {A.}~\bibnamefont
  {Brataas}},\ }\href {\doibase 10.1103/PhysRevB.95.220408} {\bibfield
  {journal} {\bibinfo  {journal} {Phys. Rev. B}\ }\textbf {\bibinfo {volume}
  {95}},\ \bibinfo {pages} {220408} (\bibinfo {year} {2017})}\BibitemShut
  {NoStop}%
\bibitem [{\citenamefont {Chotorlishvili}\ \emph {et~al.}(2015)\citenamefont
  {Chotorlishvili}, \citenamefont {Etesami}, \citenamefont {Berakdar},
  \citenamefont {Khomeriki},\ and\ \citenamefont
  {Ren}}]{chotorlishvili2015electromagnetically}%
  \BibitemOpen
  \bibfield  {author} {\bibinfo {author} {\bibfnamefont {L.}~\bibnamefont
  {Chotorlishvili}}, \bibinfo {author} {\bibfnamefont {S.}~\bibnamefont
  {Etesami}}, \bibinfo {author} {\bibfnamefont {J.}~\bibnamefont {Berakdar}},
  \bibinfo {author} {\bibfnamefont {R.}~\bibnamefont {Khomeriki}}, \ and\
  \bibinfo {author} {\bibfnamefont {J.}~\bibnamefont {Ren}},\ }\href@noop {}
  {\bibfield  {journal} {\bibinfo  {journal} {Physical Review B}\ }\textbf
  {\bibinfo {volume} {92}},\ \bibinfo {pages} {134424} (\bibinfo {year}
  {2015})}\BibitemShut {NoStop}%
\bibitem [{\citenamefont {Johansen}\ \emph {et~al.}(2018)\citenamefont
  {Johansen}, \citenamefont {Skarsv\aa{}g},\ and\ \citenamefont
  {Brataas}}]{PhysRevB.97.054423}%
  \BibitemOpen
  \bibfield  {author} {\bibinfo {author} {\bibfnamefont {O.}~\bibnamefont
  {Johansen}}, \bibinfo {author} {\bibfnamefont {H.}~\bibnamefont
  {Skarsv\aa{}g}}, \ and\ \bibinfo {author} {\bibfnamefont {A.}~\bibnamefont
  {Brataas}},\ }\href {\doibase 10.1103/PhysRevB.97.054423} {\bibfield
  {journal} {\bibinfo  {journal} {Phys. Rev. B}\ }\textbf {\bibinfo {volume}
  {97}},\ \bibinfo {pages} {054423} (\bibinfo {year} {2018})}\BibitemShut
  {NoStop}%
\bibitem [{\citenamefont {Chen}\ \emph {et~al.}(2013)\citenamefont {Chen},
  \citenamefont {Takahashi}, \citenamefont {Nakayama}, \citenamefont
  {Althammer}, \citenamefont {Goennenwein}, \citenamefont {Saitoh},\ and\
  \citenamefont {Bauer}}]{PhysRevB.87.144411}%
  \BibitemOpen
  \bibfield  {author} {\bibinfo {author} {\bibfnamefont {Y.-T.}\ \bibnamefont
  {Chen}}, \bibinfo {author} {\bibfnamefont {S.}~\bibnamefont {Takahashi}},
  \bibinfo {author} {\bibfnamefont {H.}~\bibnamefont {Nakayama}}, \bibinfo
  {author} {\bibfnamefont {M.}~\bibnamefont {Althammer}}, \bibinfo {author}
  {\bibfnamefont {S.~T.~B.}\ \bibnamefont {Goennenwein}}, \bibinfo {author}
  {\bibfnamefont {E.}~\bibnamefont {Saitoh}}, \ and\ \bibinfo {author}
  {\bibfnamefont {G.~E.~W.}\ \bibnamefont {Bauer}},\ }\href {\doibase
  10.1103/PhysRevB.87.144411} {\bibfield  {journal} {\bibinfo  {journal} {Phys.
  Rev. B}\ }\textbf {\bibinfo {volume} {87}},\ \bibinfo {pages} {144411}
  (\bibinfo {year} {2013})}\BibitemShut {NoStop}%
\bibitem [{\citenamefont {Wang}\ \emph {et~al.}(2018)\citenamefont {Wang},
  \citenamefont {Zhou}, \citenamefont {Nie}, \citenamefont {Xia},\ and\
  \citenamefont {Guo}}]{PhysRevB.97.094401}%
  \BibitemOpen
  \bibfield  {author} {\bibinfo {author} {\bibfnamefont {X.-g.}\ \bibnamefont
  {Wang}}, \bibinfo {author} {\bibfnamefont {Z.-w.}\ \bibnamefont {Zhou}},
  \bibinfo {author} {\bibfnamefont {Y.-z.}\ \bibnamefont {Nie}}, \bibinfo
  {author} {\bibfnamefont {Q.-l.}\ \bibnamefont {Xia}}, \ and\ \bibinfo
  {author} {\bibfnamefont {G.-h.}\ \bibnamefont {Guo}},\ }\href {\doibase
  10.1103/PhysRevB.97.094401} {\bibfield  {journal} {\bibinfo  {journal} {Phys.
  Rev. B}\ }\textbf {\bibinfo {volume} {97}},\ \bibinfo {pages} {094401}
  (\bibinfo {year} {2018})}\BibitemShut {NoStop}%
\bibitem [{\citenamefont {Xiao}\ \emph {et~al.}(2010)\citenamefont {Xiao},
  \citenamefont {Bauer}, \citenamefont {Uchida}, \citenamefont {Saitoh},\ and\
  \citenamefont {Maekawa}}]{PhysRevB.81.214418}%
  \BibitemOpen
  \bibfield  {author} {\bibinfo {author} {\bibfnamefont {J.}~\bibnamefont
  {Xiao}}, \bibinfo {author} {\bibfnamefont {G.~E.~W.}\ \bibnamefont {Bauer}},
  \bibinfo {author} {\bibfnamefont {K.-c.}\ \bibnamefont {Uchida}}, \bibinfo
  {author} {\bibfnamefont {E.}~\bibnamefont {Saitoh}}, \ and\ \bibinfo {author}
  {\bibfnamefont {S.}~\bibnamefont {Maekawa}},\ }\href {\doibase
  10.1103/PhysRevB.81.214418} {\bibfield  {journal} {\bibinfo  {journal} {Phys.
  Rev. B}\ }\textbf {\bibinfo {volume} {81}},\ \bibinfo {pages} {214418}
  (\bibinfo {year} {2010})}\BibitemShut {NoStop}%
\end{thebibliography}%

\end{document}